\documentclass[letterpaper]{article}

\usepackage{amsmath,amsfonts,amsthm,amssymb}
\usepackage{dsfont}
\usepackage[english]{babel}
\usepackage[left=2.0cm, top=2.5cm, bottom=2.5cm, right=2.0cm]{geometry}
\usepackage{verbatim}
\usepackage{mathrsfs}
\usepackage{bm}
\usepackage{color}
\usepackage{graphicx}
\usepackage{url}
\usepackage{caption}

\newtheorem{thm}{Theorem}[section]

\newtheorem{lem}[thm]{Lemma}

\newtheorem{prop}[thm]{Proposition}
\theoremstyle{definition}

\newtheorem{rem}[thm]{Remark}
\numberwithin{equation}{section}

\def\ed{{\,\stackrel{\mathfrak {D}}{=}\,}}
\definecolor{Red}{rgb}{0,0,0}
\newcommand{\Red}{\color{Red}}
\definecolor{DRed}{rgb}{0,0,0}

\definecolor{Green}{rgb}{0,0,0}
\newcommand{\Green}{\color{Green}}
\definecolor{Blue}{rgb}{0,0,0}

\definecolor{PaleGrey}{rgb}{0,0,0}

\title{Third-Order Short-Time Expansions for Close-to-the-Money Option Prices {Under} the CGMY Model}

\author{Jos\'{e} E. Figueroa-L\'{o}pez\thanks{{Department of Mathematics, Washington University in St. Louis, St. Louis, MO 63130, USA ({\tt figueroa@math.wustl.edu}). Research supported in part by the NSF Grants: {DMS-1561141}, DMS-1613016.}}
\and Ruoting Gong\thanks{Department of Applied Mathematics, Illinois Institute of Technology, Chicago, IL 60616, USA ({\tt rgong2@iit.edu}).}
\and Christian Houdr\'e\thanks{School of Mathematics, Georgia Institute of Technology, Atlanta, GA 30332, USA ({\tt houdre@math.gatech.edu}). Research supported in part by the grants \#246283 and \# 524678 from the Simons Foundation.}}

\date{November 21, 2017}

\begin{document}

\maketitle

\begin{abstract}
A third-order approximation for close-to-the-money European option prices under an infinite-variation CGMY L\'{e}vy model is derived, and is then extended to a model with an additional independent Brownian component. The asymptotic regime considered, in which the strike is made to converge to the spot stock price as the maturity approaches zero, is relevant in applications since the most liquid options have strikes that are close to the spot price. Our results shed new light on the connection between both the volatility of the continuous component and the jump parameters and the behavior of option prices near expiration when the strike is close to the spot price. In particular, a new type of transition phenomenon is uncovered in which the third order term exhibits two distinct asymptotic regimes depending on whether $Y\in(1,3/2)$ or $Y\in(3/2,2)$. Unlike second order approximations, the expansions herein are shown to be remarkably accurate so that they can actually be used for calibrating some model parameters. For illustration, we calibrate the volatility $\sigma$ of the Brownian component and the jump intensity $C$ of the CGMY model to actual option prices.
\end{abstract}



\section{Introduction}

Stemming in part from its importance for model testing and calibration, small-time asymptotics of option prices have received a lot of attention in recent years (see, e.g., \cite{AndersenLipton:2013}, \cite{CarrWu:2003}, \cite{FigueroaLopezForde:2012}, \cite{GatheralHsuLaurenceOuyangWang:2012}, \cite{MijatovicTankov:2013}, \cite{MuhleKarbeNutz:2011}, \cite{RoperRutkowski:2007}, \cite{Tankov:2010}, and references therein). The fact that option prices and implied volatilities exhibit sharply different behaviors under different model assumptions provides a natural tool to test the suitability of these assumptions, as already exploited by the seminal work of Carr and Wu~\cite{CarrWu:2003}. Hence, for instance, close-to-the-money implied volatilities are expected to stabilize towards a positive value (the spot volatility) near expiration under the presence of a Brownian-like component, while, in contrast, they are expected to vanish near expiration, under a pure-jump model. In both cases, the rates of convergence toward their respective steady limits are determined by the jump activity parameter $Y$, a fact that can potentially allow to assess suitable values for this parameter. Besides testing, it is important to determine what are the most important parameters driving the behavior of option prices near expiration within a class of models. For instance, within the CGMY framework in the presence of a continuous component, the most important parameter is the spot volatility, and the second most (equally) important parameters are $C$ and $Y$ . However, nothing was known related to the relevance of $G$ or $M$, before this work.

In this paper, we study the small-time behavior of close-to-the-money European call option prices
\begin{equation}\label{CallPriceDfn}
\mathbb{E}\left(\left(S_{t}-S_{0}e^{\kappa_{t}}\right)^{+}\right)=S_{0}\,\mathbb{E}\left(\left(e^{X_{t}}-e^{\kappa_{t}}\right)^{+}\right),\quad t\geq 0,
\end{equation}
where $t\to\kappa_{t}$ is a deterministic function such that $\kappa_{t}\rightarrow 0$ as $t\rightarrow 0$, and for an exponential L\'{e}vy model of the form:
\begin{equation}\label{ExpLvMdl}
S_{t}:=S_{0}e^{X_{t}},\quad\text{with }\,\,X_{t}:=L_{t}+\sigma W_{t},\quad t\geq 0.
\end{equation}
Here, $L=(L_{t})_{t\geq 0}$ is a CGMY L\'{e}vy process (cf.~\cite{CarrGemanMadanYor:2002}), while $W=(W_{t})_{t\geq 0}$ is an independent standard Brownian motion (as usual, $x^{+}:=x{\bf 1}_{\{x>0\}}$ and $x^{-}:=x{\bf 1}_{\{x<0\}}$ denote the positive and negative parts of a real $x$).

The asymptotic behavior of (\ref{CallPriceDfn}) is known to change radically depending on whether the parameter $Y$ of the process $L$ is smaller or larger than $1$ (cf.~\cite{Tankov:2010}). We focus here on the latter case, which arguably is more relevant for financial applications, in light of some recent empirical evidence based on high-frequency data supporting this assumption (cf.~\cite{AitSahaliaJacod:2009}, \cite{Belomestny:2010}, and the references therein). In the pure-jump CGMY case ($\sigma=0$), {it is known} (cf.~\cite{FigueroaGongHoudre:2012(2)}) that the short-time second-order asymptotic behavior of the ATM call option price is of the form
\begin{align*}
\frac{1}{S_{0}}\mathbb{E}\left(\left(S_{t}-S_{0}\right)^{+}\right)=d_{1}t^{\frac{1}{Y}}+d_{2}t+o(t),\quad t\rightarrow 0,
\end{align*}
while in the case of a non-zero independent Brownian component ($\sigma\neq 0$),
\begin{equation}\label{ExpAsymBehCGMYBM}
\frac{1}{S_{0}}\mathbb{E}\left(\left(S_{t}-S_{0}\right)^{+}\right)=d_{1}t^{\frac{1}{2}}+d_{2}t^{\frac{3-Y}{2}}+o\left(t^{\frac{3-Y}{2}}\right),\quad t\rightarrow 0,
\end{equation}
for (different) constants $d_{1}$ and $d_{2}$, which are explicitly given in the sequel. For extensions of these results to a more general class of processes, we refer the reader to~\cite{FigueroaLopezGongHoudre:2014} and~\cite{FigueroaLopezOlafsson:2014}.

In this {\Red paper}, we derive the third-order asymptotic behavior for the close-to-the-money option prices (\ref{CallPriceDfn}) in the CGMY model both with and without an independent Brownian component, when the log-moneyness $\kappa_{t}$ converges to $0$ at a suitable rate, as the maturity $t$ goes to $0$. Our motivations for considering these expansions are twofold. First, though being a significant improvement over the first-order expansion, in some cases the second-order expansion might not be that accurate unless $t$ is extremely small (see the numerical examples provided in~\cite[Section 6]{FigueroaLopezGongHoudre:2014} {and also in Section \ref{NumExples} herein)}. This is particularly true in the presence of an independent Brownian component. As shown in the sequel, the third-order expansions, derived here, can {dramatically} improve the approximation's accuracy, {even for maturities as large as a few years (see, e.g., {\Red the} left panel of Figure \ref{FgATMP} in Section \ref{NumExples}). As shown herein, this improvement in accuracy allows to provide adequate calibration of some model parameters such as the volatility $\sigma$ of the Brownian component and the jump intensity $C$ of the CGMY model}. Second, the expansions developed here shed a new light on the effects of both the volatility of the continuous component and the jump parameters in the behavior of option prices, near expiration, when the strike is close to the spot price. In particular, in the same way as the asymptotic behavior of the leading term substantially changes when $Y$ transitions at $1$, we uncover a similar phenomenon for the third-order term, but this time when $Y$ transitions at $3/2$. This identifies the value of $Y=3/2$ as another transition point for the asymptotic behavior of ATM option prices.

As in~\cite{FigueroaLopezGongHoudre:2014}, an important ingredient in our approach is a change of probability measure, under which $(L_{t})_{t\geq 0}$ becomes a stable L\'{e}vy process, enabling us to exploit high-order asymptotics for the transition densities of such processes. However, the extension from the second-order to the third-order asymptotics for option prices is quite intricate and requires the development of some new techniques beyond those used in~\cite{FigueroaLopezGongHoudre:2014}. For instance, as it turns out, an important step in obtaining the asymptotic expansion in the presence of an independent Brownian component is to determine the short-time asymptotics of the following quantity:
\begin{align}\label{NdedAsymp}
R^{(k)}_{t}:=\int_{0}^{\infty}\mathbb{E}\left(\left(\sigma W_{1}\right)^{k}{\bf 1}_{\left\{0\leq\sigma W_{1}\leq tz\right\}}\right)z^{1-k}\left(p_{Z}(z)-Cz^{-Y-1}\right)dz,\quad\text{for }\,k=0,1,
\end{align}
where $p_{Z}$ is the density of a symmetric stable random variable $Z$ with a L\'{e}vy density $C|x|^{-Y-1}$ so that
\begin{align}\label{Asympp0}
p_{Z}(z)=Cz^{-Y-1}+C'z^{-2Y-1}+o\left(z^{-2Y-1}\right),\quad z\rightarrow\infty,
\end{align}
for an appropriate constant $C'$ (see (\ref{Asydenpz}) below for details). A natural idea to analyze (\ref{NdedAsymp}) is then to plug (\ref{Asympp0}) in (\ref{NdedAsymp}) and change variables to $u=tz$ to get
\begin{align*}
R^{(k)}_{t}\sim C'\,t^{2Y+k-1}\int_{0}^{\infty}\mathbb{E}\left(\left(\sigma W_{1}\right)^{k}{\bf 1}_{\left\{0\leq\sigma W_{1}\leq u\right\}}\right)u^{-2Y-k}du=-C'\,\frac{\sigma^{1-2Y}}{2(2Y+k-1)}\,t^{2Y+k-1}\mathbb{E}\left(\left|W_{1}\right|^{1-2Y}\right),
\end{align*}
where, in the last equality, Fubini's theorem and the symmetry of $W_{1}$ were used. However, when $Y>1$, $1-2Y<-1$, and the last expectation is infinite, which shows that the above heuristic argument is false. Instead, in this work, we make use of Fourier analysis techniques for {tempered distributions} to handle (\ref{NdedAsymp}). This method, interesting on its own, is new and differs from the arguments developed in our earlier work in~\cite{FigueroaLopezGongHoudre:2014}.

The remaining of the paper is organized as follows. Section 2 contains preliminary results on the CGMY model, some probability measure transformations, and asymptotic results for stable L\'{e}vy processes, which will be needed throughout the paper. Section 3 establishes the third-order asymptotics for close-to-the-money call option prices, as well as the corresponding Black-Scholes implied volatilities, under both the pure-jump CGMY model ($\sigma=0$) and the CGMY model with an independent Brownian component ($\sigma\neq 0$). Some numerical examples are also provided in this section to illustrate the high performance of our asymptotic expansions, {together with an actual calibration exercise with real option data}. The proofs of our main results are deferred to the Appendix.

\section{Setup and Preliminary Results}\label{Sec:Setup}

Throughout, $W:=(W_{t})_{t\geq 0}$ and $L:=(L_{t})_{t\geq 0}$ respectively stand for a standard Brownian motion and a pure-jump CGMY L\'{e}vy process independent of each other (cf.~\cite{CarrGemanMadanYor:2002}) defined on a complete filtered probability space $(\Omega,\mathcal{F},(\mathcal{F}_{t})_{t\geq 0},\mathbb{P})$. As usual, we denote the parameters of $L$ by $C,\,G,\,M>0$, and $Y\in(0,2)$ so that the L\'evy measure of $L$ is given by $\nu(dx)=C|x|^{-Y-1}(e^{-Mx}\,{\bf 1}_{\{x>0\}}+e^{Gx}\,{\bf 1}_{\{x<0\}})dx$. Hereafter, we assume $Y\in(1,2)$, $M>1$, zero interest rate, and that $\mathbb{P}$ is a martingale measure for the exponential L\'{e}vy model (\ref{ExpLvMdl}) with the log-return process $X_{t}:=L_{t}+\sigma W_{t}$, $t\geq 0$. The following notation is also used in what follows:
\begin{align*}
M^{*}=M-1,\quad G^{*}=G+1,\quad \varphi(x):=M^{*}x\,{\bf 1}_{\{x>0\}}-G^{*}x\,{\bf 1}_{\{x<0\}},\quad\nu^{*}(dx)=e^{x}\nu(dx).
\end{align*}
We will make use of two density transformations of the L\'evy process (cf.~\cite[Definition 33.4]{Sato:1999}). Hereafter, $\mathbb{P}^{*}$ and $\widetilde{\mathbb{P}}$ are probability measures on $(\Omega,\mathcal{F})$ such that, for any {$t\geq 0$,
\begin{align}\label{DSMBoth}
\ln\left(\frac{d\mathbb{P}^{*}|_{\mathcal{F}_{t}}}{d\mathbb{P}|_{\mathcal{F}_{t}}}\right)=X_{t},\quad\quad\ln\left(\frac{d\widetilde{\mathbb{P}}\big|_{\mathcal{F}_{t}}}{d\mathbb{P}^{*}\big|_{\mathcal{F}_{t}}}\right)=U_{t}:=\lim_{\epsilon\rightarrow 0}\left(\int_{0}^{t}\int_{|x|>\varepsilon}\varphi(x)N(ds,dx)-t\int_{|x|>\varepsilon}(e^{\varphi(x)}-1)\nu^{*}(dx)\right),
\end{align}
where $N(dt,dx):=\#\{(s,\Delta X_{s})\in dt\times dx\}$ is the jump measure of $(X_{t})_{t\geq 0}$}. Throughout, $\mathbb{E}^{*}$ and $\widetilde{\mathbb{E}}$ denote the expectations under $\mathbb{P}^{*}$ and $\widetilde{\mathbb{P}}$, respectively.

From the density transformation and the L\'{e}vy-It\^{o} decomposition of a L\'{e}vy process (cf.~\cite[Theorems 19.2 and Theorem 33.1]{Sato:1999}), $X_{t}=L_{t}^{*}+\sigma W_{t}^{*}$, {$t\geq 0$,} where, under $\mathbb{P}^{*}$, $(W_{t}^{*})_{t\geq 0}$ is again a Wiener process while $(L_{t}^{*})_{t\geq 0}$ is still a CGMY {L\'{e}vy} process, independent of $W^{*}$, but with parameters $C$, $Y$, $M=M^{*}$ and $G=G^{*}$. The L\'{e}vy triplet of $(X_{t})_{t\geq 0}$ under $\mathbb{P}^{*}$ is given by $(b^{*},(\sigma^{*})^{2},\nu^{*})$ with $\sigma^{*}:=\sigma$, $\nu^{*}(dx)=C|x|^{-Y-1}(e^{-M^{*}x}\,{\bf 1}_{\{x>0\}}+e^{G^{*}x}\,{\bf 1}_{\{x<0\}})dx$, and
\begin{align*}
b^{*}:=-C\Gamma(-Y)\left[\left(M^{*}\right)^{Y}+\left(G^{*}\right)^{Y}-M^{Y}-G^{Y}\right]+\frac{\sigma^{2}}{2}-\int_{|x|>1}x\,\nu^{*}(dx)-CY\Gamma(-Y)\left[\left(M^{*}\right)^{Y-1}-\left(G^{*}\right)^{Y-1}\right].
\end{align*}
Under the measure $\widetilde{\mathbb{P}}$, the process $(L_{t}^{*})_{t\geq 0}$ becomes a stable L\'{e}vy process while $(W_{t}^{*})_{t\geq 0}$ remains a Wiener process independent of $L^{*}$. Concretely, setting $\tilde{\nu}(dx):=C|x|^{-Y-1}dx$ and $\tilde{b}=b^{*}+\int_{|x|\leq 1}x\left(\tilde{\nu}-\nu^{*}\right)(dx)$, $(X_{t})_{t\geq 0}$ is a L\'{e}vy process with L\'{e}vy triplet $(\tilde{b},\sigma^{2},\tilde\nu)$, under $\widetilde{\mathbb{P}}$. In particular, letting
\begin{align}\label{Cent}
\tilde{\gamma}:=\widetilde{\mathbb{E}}\left(X_{1}\right)=-C\Gamma(-Y)\left[(M-1)^{Y}+(G+1)^{Y}-M^{Y}-G^{Y}\right]+\frac{\sigma^{2}}{2},
\end{align}
the centered process $Z_{t}:=L_{t}^{*}-t\tilde{\gamma}$ is symmetric and strictly $Y$-stable under $\widetilde{\mathbb{P}}$.

It will be convenient to express the process $(U_{t})_{t\geq 0}$ defined in (\ref{DSMBoth}) in terms of the compensated measure $\bar{N}(dt,dx):=N(dt,dx)-\tilde{\nu}(dx)dt$ (under $\widetilde{\mathbb{P}}$), namely,
\begin{align}\label{DcmLL}
U_{t}=M^{*}\bar{U}_{t}^{(p)}-G^{*}\bar{U}_{t}^{(n)}+\eta t=:\widetilde{U}_{t}+\eta t,\quad t\geq 0,
\end{align}
where
\begin{align}\label{eta}
\bar{U}^{(p)}_{t}:=\int_{0}^{t}\int_{0}^{\infty}x\,\bar{N}(ds,dx),\quad\bar{U}^{(n)}_{t}:=\int_{0}^{t}\int_{-\infty}^{0}x\,\bar{N}(ds,dx),\quad\eta:=C\Gamma(-Y)\left[(M-1)^{Y}+(G+1)^{Y}\right].
\end{align}
Note that, under $\widetilde{\mathbb{P}}$, $(\bar{U}_{t}^{(p)})_{t\geq 0}$ and $(-\bar{U}_{t}^{(n)})_{t\geq 0}$ are independent and identically distributed one-sided $Y$-stable processes with scale, skewness, and location parameters given by $C|\cos(\pi Y/2)|\Gamma(-Y)$, $1$, and $0$, respectively. Finally, let us further note the following decomposition of the process $X$ in terms of the {processes previously defined}:
\begin{align}\label{RX}
X_{t}=Z_{t}+t\tilde\gamma+\sigma W_{t}^{*}=\bar{U}^{(p)}_{t}+\bar{U}^{(n)}_{t}+t\tilde{\gamma}+\sigma W_{t}^{*},\quad t\geq 0.
\end{align}

To conclude this section, we recall some well-known results on the transition densities of stable processes. The following second-order approximation of the density $p_{U}(1,x)$ of $\bar{U}_{1}^{(p)}\ed -\bar{U}_{1}^{(n)}$ is well-known (cf. (14.34) in~\cite{Sato:1999}):
\begin{align}\label{eq:2ndAsyStableDen}
p_{U}(1,x)=Cx^{-Y-1}-\frac{C^{2}}{2\pi}\sin(2\pi Y)\Gamma(2Y+1)\Gamma^{2}(-Y)x^{-2Y-1}+o\left(x^{-2Y-1}\right),\quad x\rightarrow\infty.
\end{align}
In particular,
\begin{align}\label{eq:2ndAsyStableTailDist}
\widetilde{\mathbb{P}}\left(\bar{U}_{1}^{(p)}\geq x\right)=\widetilde{\mathbb{P}}\left(-\bar{U}_{1}^{(n)}\geq x\right)=\frac{C}{Y}x^{-Y}-\frac{C^{2}}{2\pi}\sin(2\pi Y)\Gamma(2Y)\Gamma^{2}(-Y)x^{-2Y}+o\left(x^{-2Y}\right),\quad x\rightarrow\infty.
\end{align}
The following result sharpens (\ref{eq:2ndAsyStableDen}) and (\ref{eq:2ndAsyStableTailDist}). The proof of (\ref{UIn1Gen}-i) is given in~\cite{FigueroaLopezGongHoudre:2014}, while (\ref{UIn1Gen}-ii) is presented in the Appendix \ref{proofB}.
\begin{lem}\label{Bnd1TailSt}
There exist constants $0<{K}_{1},{K}_{2}<\infty$ such that, for \emph{any} $x>0$,
\begin{align}\label{UIn1Gen}
{\rm (i)}\,\,\widetilde{\mathbb{P}}\!\left(\bar{U}_{1}^{(p)}\!\!\geq\!x\right)\!=\!\widetilde{\mathbb{P}}\!\left(-\bar{U}_{1}^{(n)}\!\!\geq\!x\right)\!\leq\!K_{1}x^{-Y},\,\,\,
{\rm (ii)}\,\left|\,\widetilde{\mathbb{P}}\!\left(\bar{U}_{1}^{(p)}\!\!\geq\!x\right)\!-\!\frac{C}{Y}x^{-Y}\right|\!=\!\left|\widetilde{\mathbb{P}}\!\left(-\bar{U}_{1}^{(n)}\!\!\geq\!x\right)\!-\!\frac{C}{Y}x^{-Y}\right|\!\leq\!K_{2}x^{-2Y}.
\end{align}
\end{lem}

Similarly, the tail distribution and the probability density of $Z_{1}$, hereafter denoted by $p_{Z}(1,z)$, admit the following asymptotic behavior\footnote{In terms of the parametrization in~\cite[Definition 14.16]{Sato:1999}, $(\alpha,\beta,\tau,c)$ of $Z_{1}$ therein is $(Y,0,0,2C|\cos(\pi Y/2)|\Gamma(-Y))$.} (cf. (14.34) in \cite{Sato:1999}):
\begin{align}
\widetilde{\mathbb{P}}(Z_{1}\geq z)&=\frac{C}{Y}z^{-Y}-\frac{C^{2}}{\pi Y}\sin(\pi Y)\cos^{2}\left(\frac{\pi Y}{2}\right)\Gamma(2Y+1)\Gamma^{2}(-Y)z^{-2Y}+o\left(z^{-2Y}\right),\quad z\rightarrow\infty,\nonumber\\
\label{Asydenpz} p_{Z}(1,z)&=Cz^{-Y-1}-\frac{2C^{2}}{\pi}\sin(\pi Y)\cos^{2}\left(\frac{\pi Y}{2}\right)\Gamma(2Y+1)\Gamma^{2}(-Y)z^{-2Y-1}+o\left(z^{-2Y-1}\right),\quad z\rightarrow\infty.
\end{align}

\section{The Main Results}

In this section, we give the third-order asymptotic behavior for near at-the-money call option prices {\Red and implied volatilities} in both the pure-jump ($\sigma=0$) and the mixed ($\sigma\neq 0$) models. {We first consider the expansion for the latter case since {\Red it} is more explicit and of greater use for financial application in view of some empirical evidence, based on high-frequency data, which tends to support a mixed model over either a pure-jump or a purely continuous one (cf.~\cite{AitSahaliaJacod:2010}). The {\Red results} for the pure-jump case {\Red are} given at the end of the section.} The proofs of the main results are deferred to Appendix \ref{proofA}.

{{\Red For} a mixed CGMY model with the addition of an independent Brownian component,} it was shown in~\cite[Section 5]{FigueroaLopezGongHoudre:2014} that, the second-order correction term for the ATM European call option price is given by (\ref{ExpAsymBehCGMYBM}) with
\begin{align}\label{PrcDefn2ndTrm}
d_{1}:=\mathbb{E}^{*}\left(\left(\sigma W_{1}^{*}\right)^{+}\right)=\frac{\sigma}{\sqrt{2\pi}},\quad d_{2}:=\frac{C\sigma^{1-Y}}{Y(Y-1)}\widetilde{\mathbb{E}}\left(\left|W_{1}^{*}\right|^{1-Y}\right)=\frac{C2^{\frac{1-Y}{2}}\sigma^{1-Y}}{\sqrt{\pi}Y(Y-1)}\Gamma\left(1-\frac{Y}{2}\right).
\end{align}
As observed from these expressions, the first-order term only synthesizes the information about the continuous volatility parameter $\sigma$, while the second-order term also incorporates the information on the degree of jump activity $Y$ and the overall jump-intensity parameter $C$. However, these two terms do not reflect the relative intensities of the negative or positive jumps (as given by the parameters $G$ and $M$, respectively). {{\Red Moreover}, as we shall see in the numerical experiments of Section \ref{NumExples}, the second order expansion is still quite imprecise for an adequate calibration of the parameters $\sigma$ and $C$, even if we {\Red fix} the value of $Y$. These facts suggest} the need for a higher-order approximation as described in the forthcoming theorem.

As mentioned {in the introduction}, the following preliminary result will play a crucial role in the proof of Theorem \ref{thm:ThirdOrderAsyCGMYB}.
\begin{lem}\label{thm:ThirdOrderAsyCGMYBLemma}
Let $R^{(k)}_{t}$ be as in (\ref{NdedAsymp}). Then, $R^{(k)}_{t}\sim t^{2Y+k-1}E^{(k)}$, where
\begin{align}\label{DfnE01}
E^{(0)}=-\frac{2C^{2}Y\cos^{2}\left(\frac{\pi Y}{2}\right)\Gamma^{2}(-Y)}{\sqrt{2\pi}\sigma^{2Y-1}}\mathbb{E}\left(\left|W_{1}\right|^{2Y-2}\right),\quad E^{(1)}=-\frac{C^{2}(2Y-1)\cos^{2}\left(\frac{\pi Y}{2}\right)\Gamma^{2}(-Y)}{\sqrt{2\pi}\sigma^{2Y-1}}\mathbb{E}\left(\left|W_{1}\right|^{2Y-2}\right).
\end{align}
\end{lem}
\begin{thm}\label{thm:ThirdOrderAsyCGMYB}
Let $t\mapsto\kappa_{t}$, $t\geq 0$, be a deterministic function such that $\kappa_{t}=o(1)$ as $t\to 0$. Let also
\begin{align}\label{eq:3rdCoefGenCGMY1}
d_{31}&:={\frac{C\Gamma(-Y)}{2}\left[(M-1)^{Y}-M^{Y}-(G+1)^{Y}+G^{Y}\right]},\\
\label{eq:3rdCoefGenCGMY2} d_{32}&:=-\frac{1}{\pi}\sigma^{1-2Y}C^{2}\cos^{2}\left(\frac{\pi Y}{2}\right)\Gamma^{2}(-Y)2^{Y-\frac{1}{2}}\Gamma\left(Y-\frac{1}{2}\right),\\
c_{\kappa,\sigma}(t)&:=\kappa_{t}\int_{0}^{1}\mathbb{P}\left(\sigma W_{t}\geq{}\kappa_{t}w\right)dw.\nonumber
\end{align}
Then, under the exponential CGMY model (\ref{ExpLvMdl}) with an independent Brownian component,
\begin{equation}\label{3rdAsyCGMYB1}
{\frac{e^{-\kappa_{t}}}{S_{0}}}\mathbb{E}\left(\left(S_{t}-S_{0}e^{\kappa_{t}}\right)^{+}\right) {+c_{\kappa,\sigma}(t)}=d_{1}t^{\frac{1}{2}}+d_{2}t^{\frac{3-Y}{2}}+d_{31}t+d_{32}t^{\frac{5}{2}-Y}+{o(\kappa_{t})}+o(t)+o\left(t^{\frac{5}{2}-Y}\right),\quad t\rightarrow 0,
\end{equation}
where $d_{1}$ and $d_{2}$ are given as in (\ref{PrcDefn2ndTrm}) {and the terms $o(t)$ and $o\left(t^{5/2-Y}\right)$ do not depend on $\kappa_{t}$}.
\end{thm}
\begin{rem}
{The form of the asymptotic expansion (\ref{3rdAsyCGMYB1}), which {\Red might, at first, appear unconventional}, is chosen with the aim to partially disentangle the effect of the log-moneyness $\kappa_{t}$, whose value is actually observed, from the option price. If we further assume that $\kappa_{t}=o(\sqrt{t})$, {\Red the expansion becomes}
\begin{align*}	 \frac{e^{-\kappa_{t}}}{S_{0}}\mathbb{E}\left(\left(S_{t}-S_{0}e^{\kappa_{t}}\right)^{+}\right)+\frac{\kappa_{t}}{2}=d_{1}t^{\frac{1}{2}}+d_{2}t^{\frac{3-Y}{2}}+d_{31}t+d_{32}t^{\frac{5}{2}-Y}+o(t)+o\left(t^{\frac{5}{2}-Y}\right)+o(\kappa_{t}),\quad t\rightarrow 0.
\end{align*}
{\Red The} quantity appearing on the {\Red left-hand} side of the above {\Green equation, {\Red is then called the} log-moneyness adjusted price}. {\Red This quantity} can be easily computed since $\kappa_{t}$ is known. If $Y\in(1,3/2)$ and $\kappa_{t}=O(t)$, the third-order term of the log-moneyness adjusted price is $d_{31}t$; if $Y\in(3/2,2)$ and $\kappa_{t}=O(t^{5/2-Y})$, the third-order term is $d_{32}t^{\frac{5}{2}-Y}$; finally, if $Y=3/2$ and $\kappa_{t}=O(t)$, the third-order {\Red term} is $(d_{31}+d_{32})t$.}
\end{rem}

Our next proposition gives the small-time asymptotic behavior for the {\Red close-to-the-money} Black-Scholes implied volatility, {\Red hereafter denoted} by $\hat{\sigma}$, corresponding to the option prices of Theorem \ref{thm:ThirdOrderAsyCGMYB}. The proof is similar to that of~\cite[Corollary 4.3]{FigueroaLopezGongHoudre:2014} and is therefore omitted.
\begin{prop}\label{Prop:AsyIVCGMYB}
Let {\Red $d_{1}$,} $d_{2}$, $d_{31}$, and $d_{32}$ be respectively given by (\ref{PrcDefn2ndTrm}), (\ref{eq:3rdCoefGenCGMY1}), and (\ref{eq:3rdCoefGenCGMY2}), and let $d_{3}=d_{31}{\bf 1}_{\{Y\leq 3/2\}}+d_{32}{\bf 1}_{\{Y\geq 3/2\}}$. Suppose that the log-moneyness $\kappa_{t}$ is such that {\Red $\kappa_{t}=o(\sqrt{t})$, as $t\rightarrow 0$}.
Then, under the exponential CGMY model (\ref{ExpLvMdl}) with an independent Brownian component, as $t\rightarrow 0$,
\begin{align}\label{ExpSigmaImBM}
\frac{1}{\sqrt{2\pi}}\hat\sigma(t)+{\frac{\kappa_{t}}{2\sqrt{t}}}=\left\{\begin{array}{ll}
{\Red d_{1}}+d_{2}t^{1-\frac{Y}{2}}+d_{3}t^{\frac{1}{2}}+o\left(t^{\frac{1}{2}}\right),&\text{if }\,1<Y\leq 3/2,\vspace{0.2 cm}\\
{\Red d_{1}}+d_{2}t^{1-\frac{Y}{2}}+d_{3}t^{2-Y}+o\left(t^{2-Y}\right),&\text{if }\,3/2<Y<2.\end{array}\right.
\end{align}
\end{prop}
{Note that the coefficients of the log-moneyness expansion are not needed in (\ref{ExpSigmaImBM}) and the left-hand side of that equation, which {\Green we could call} the log-moneyness adjusted implied volatility, can actually be computed since we typically {\Green know} the value of $\kappa_{t}$.}

We now analyze the case of a pure-jump CGMY model.
\begin{thm}\label{thm:ThirdOrderAsyPureCGMY}
Let $t\to\kappa_{t}$ be a deterministic function such that {\Red $\kappa_{t}=o(1)$ as $t\rightarrow 0$. Let also
\begin{align}\label{eq:1stCoefPureCGMY} d_{1}&:=\widetilde{\mathbb{E}}\left(Z_{1}^{+}\right)=\frac{1}{\pi}\Gamma\left(1-\frac{1}{Y}\right)\left(2C\Gamma(-Y)\left|\cos\left(\frac{\pi Y}{2}\right)\right|\right)^{\frac{1}{Y}},\\
\label{eq:2ndCoefPureCGMY} d_{2}&:=\frac{C\Gamma(-Y)}{2}\left[(M-1)^{Y}-M^{Y}-(G+1)^{Y}+G^{Y}\right],\\
\label{eq:3rdCoefPureCGMY1} d_{31}&:=\frac{\tilde{\gamma}^{2}}{2}p_{Z}(1,0)= \frac{\tilde{\gamma}^{2}}{2\pi}\Gamma\left(1+\frac{1}{Y}\right)\left(-2C\Gamma(-Y)\cos\left(\frac{\pi Y}{2}\right)\right)^{-\frac{1}{Y}},\\
\label{eq:3rdCoefPureCGMY2} d_{32}&:=-\frac{1}{2}\widetilde{\mathbb{E}}\left(\left(Z_{1}^{+}+\widetilde{U}_{1}\right)^{2}{\bf 1}_{\{Z_{1}^{+}+\widetilde{U}_{1}\leq 0\}}\right)-\int_{0}^{\infty}w\left(\widetilde{\mathbb{P}}\left(Z_{1}^{+}+\widetilde{U}_{1}\geq w\right)-\frac{CM^{Y}}{Yw^{Y}}-\frac{C(G+1)^{Y}}{Yw^{Y}}\right)dw,\\
\widetilde{c}_{\kappa}(t)&:=\kappa_{t}\int_{0}^{1}\widetilde{\mathbb{P}}(Z_{t}\geq\kappa_{t}w)\,dw.\nonumber
\end{align}
Then}, under the exponential CGMY model (\ref{ExpLvMdl}) without a Brownian component,
\begin{align}\label{eq:3rdExpPureCGMY1}
{\Red \frac{e^{-\kappa_{t}}}{S_{0}}}\mathbb{E}\left(\left(S_{t}-S_{0}e^{\kappa_{t}}\right)^{+}\right)+{\Red \widetilde{c}_{\kappa}(t)}=d_{1}t^{\frac{1}{Y}}+d_{2}t+d_{31}t^{2-\frac{1}{Y}}+d_{32}t^{\frac{2}{Y}}+{\Red o(\kappa_{t})}+o\left(t^{2-\frac{1}{Y}}\right)+o\left(t^{\frac{2}{Y}}\right),\quad t\rightarrow 0,
\end{align}
where {\Red the terms $o(t^{2-1/Y})$ and $o(t^{2/Y})$ do not depend on $\kappa_{t}$.}
\end{thm}
{\Red
\begin{rem}
In the pure-jump case, if we further assume that $\kappa_{t}=o(t^{1/Y})$, then (\ref{eq:3rdExpPureCGMY1}) becomes
\begin{align*}
\frac{e^{-\kappa_{t}}}{S_{0}}\mathbb{E}\left(\left(S_{t}-S_{0}e^{\kappa_{t}}\right)^{+}\right)+\frac{\kappa_{t}}{2}=d_{1}t^{\frac{1}{Y}}+d_{2}t+d_{31}t^{2-\frac{1}{Y}}+d_{32}t^{\frac{2}{Y}}+o(\kappa_{t})+o\left(t^{2-\frac{1}{Y}}\right)+o\left(t^{\frac{2}{Y}}\right),\quad t\rightarrow 0.
\end{align*}
In particular, if $Y\in(1,3/2)$ and $\kappa_{t}=O(t^{2-1/Y})$, the third-order term of the log-moneyness adjusted price is $d_{31}t^{2-1/Y}$; if $Y\in(3/2,2)$ and $\kappa_{t}=O(t^{2/Y})$, the third-order term is $d_{32}t^{2/Y}$; and finally, if $Y=3/2$ and $\kappa_{t}=O(t^{4/3})$, the third-order term is $(d_{31}+d_{32})t^{4/3}$.
\end{rem}
We conclude the section by stating the following small-time asymptotic expansion of the close-to-the-money} Black-Scholes implied volatility, {\Red denoted again} by $\hat{\sigma}$, {\Red corresponding to the option prices of Theorem \ref{thm:ThirdOrderAsyPureCGMY}}. The proof is similar to that of~\cite[Corollary 3.7]{FigueroaLopezGongHoudre:2014} and is therefore omitted.
\begin{prop}\label{AsyIVPCGMY}
Let $d_{1}$, $d_{2}$, $d_{31}$, and $d_{32}$ be respectively given as in (\ref{eq:1stCoefPureCGMY})$-$(\ref{eq:3rdCoefPureCGMY2}), and let $d_{3}=d_{31}{\bf 1}_{\{Y\leq 3/2\}}+d_{32}{\bf 1}_{\{Y\geq 3/2\}}$. {\Red Suppose that the log-moneyness $\kappa_{t}$ is such that $\kappa_{t}=o(t^{1/Y})$, as $t\rightarrow 0$.
Then}, under the exponential CGMY model (\ref{ExpLvMdl}) without a Brownian component, as $t\rightarrow 0$,
\begin{align*}
\frac{1}{\sqrt{2\pi}}\hat\sigma(t){\Red +\frac{\kappa_{t}}{2\sqrt{t}}}=\left\{\begin{array}{ll} d_{1}t^{\frac{1}{Y}-\frac{1}{2}}+d_{2}t^{\frac{1}{2}}+d_{3}t^{\frac{3}{2}-\frac{1}{Y}}+o\left(t^{\frac{3}{2}-\frac{1}{Y}}\right),&\text{if }\,1<Y\leq 3/2,\vspace{0.2 cm}\\
d_{1}t^{\frac{1}{Y}-\frac{1}{2}}+d_{2}t^{\frac{1}{2}}+d_{3}t^{\frac{2}{Y}-\frac{1}{2}}+o\left(t^{\frac{2}{Y}-\frac{1}{2}}\right), &\text{if }\,3/2<Y<2.\end{array}\right.
\end{align*}
\end{prop}
\section{Numerical Examples}\label{NumExples}

\subsection{Performance of Approximations}

This section is devoted to assess the performance of the previous approximations. For simplicity, we assume $S_{0}=1$ and zero interest rate throughout this section. There are two popular numerical methods to evaluate the option prices of parametric L\'evy models: Inverse Fourier Transform (IFT) and Monte Carlo (MC) Methods. As illustrated in~\cite{FigueroaLopezGongHoudre:2014}, the IFT method is less accurate than MC method when computing close-to-the-money option prices with short maturities. For instance, consider the IFT method described in \cite[Section 11.1.3]{ContTankov:2004}), which is based on the formula
\begin{align}\label{IntgIFM}
z_{t}(\kappa):=C(\kappa)-C_{BS}^{\Sigma}(\kappa)=\frac{1}{2\pi}\int_{-\infty}^{\infty}e^{-iv\kappa}\frac{\varphi_{t}(v-i)-\varphi_{t}^{BS,\Sigma}(v-i)}{iv\left(1+iv\right)}\,dv=:\frac{1}{2\pi}\int_{-\infty}^{\infty}e^{-iv\kappa}\zeta_{t}(v)\,dv,
\end{align}
where $C(\kappa)$ denotes the call option price at the log-moneyness $\kappa=\ln K$ that we wish to compute and $C_{BS}^{\Sigma}(\kappa)$ denotes the corresponding call option price at the log-moneyness $\kappa=\ln K$. Above, $\varphi_{t}^{BS,\Sigma}=\exp\Big(-\frac{\Sigma^{2}t}{2}\left(v^{2}+iv\right)\Big)$ denotes the characteristic function corresponding to the Black-Scholes model with the volatility $\Sigma$ and  $\varphi_{t}$ is the characteristic function of the log-return, under the mixed CGMY model with an independent Brownian component. To explain where the issues in applying (\ref{IntgIFM}) come from, let us note that, for a close-to-the-money regime, where $\kappa$ is close to zero, the integrand in (\ref{IntgIFM}) approximately reduces to $\zeta_{t}$, which is not easy to integrate numerically for small $t$, since in that case $\varphi_{t}$ and $\varphi_{t}^{BS,\Sigma}$ are quite flat in a large domain of the integration variable $v$ (see~\cite[Section 10]{FigueroaLopezGongHoudre:2014} for numerical results using Simpson's rule).

In this work, we apply a MC method to compute the option prices under a CGMY model. This is based on the option price representation under the probability measure $\widetilde{\mathbb{P}}$ (see also~\cite[Section 6.1]{FigueroaLopezGongHoudre:2014}). Using (\ref{DcmLL})-(\ref{RX}), we have
\begin{align*}
\mathbb{E}\left[\left(e^{X_{t}}-e^{\kappa_{t}}\right)^{+}\right]&=\mathbb{E}^{*}\left[e^{-X_{t}}\left(e^{X_{t}}-e^{\kappa_{t}}\right)^{+}\right]=\widetilde{\mathbb{E}}\left[e^{-U_{t}}\left(1-e^{\kappa_{t}-X_{t}}\right)^{+}\right]\\
&=\widetilde{\mathbb{E}}\left[e^{-M^{*}\bar{U}_{t}^{(p)}+G^{*}\bar{U}_{t}^{(n)}-\eta t}\left(1-e^{\kappa_{t}-\bar{U}_{t}^{(p)}-\bar{U}_{t}^{(n)}-\tilde{\gamma}t-\sigma W_{t}^{*}}\right)^{+}\right],
\end{align*}
which can be easily computed by the MC method using the fact that, under $\widetilde{\mathbb{P}}$, $\bar{U}_{t}^{(p)}$ and $-\bar{U}_{t}^{(n)}$ are independent $Y$-stable random variables with scale, skewness and location parameters {$(tC|\cos(\pi Y/2)|\Gamma(-Y))^{1/Y}$}, $1$ and $0$, respectively. We use the simulation method of~\cite{ChambersMallowsStuck:1976} to generate the stable random variables $\bar{U}_{t}^{(p)}$ and $\bar{U}_{t}^{(n)}$.

Our parameter settings are motivated by the studies in~\cite{AndersenLipton:2013} and~\cite{Schoutens:2003}. Concretely, in~\cite{AndersenLipton:2013}, a mixed exponential L\'{e}vy model with L\'{e}vy measure
\begin{align*}
\nu(dx)=\left(\frac{C_{+}e^{-Mx}}{x^{1+Y}}\,{\bf 1}_{\{x>0\}}+\frac{C_{-}e^{Gx}}{|x|^{1+Y}}\,{\bf 1}_{\{x<0\}}\right)dx,
\end{align*}
was considered. The calibrated parameters were given as follows (see Table 5 therein) for two different stocks:
\begin{align}
C_{+}&=0.0069,\quad C_{-}=0.0063,\quad G=0.41,\quad M=1.93,\quad Y=1.5,\quad\sigma=0,\nonumber\\
\label{val2}
C_{+}&=0.0028,\quad C_{-}=0.0025,\quad G=0.41,\quad M=1.93,\quad Y=1.5,\quad\sigma=0.1.
\end{align}
In the sequel, we simply take $C:=(C_{+}+C_{-})/2$. In~\cite{Schoutens:2003}, the CGMY model was considered, and the calibrated parameters were given as (see Table 6.3 therein):
\begin{align*}
C=0.0244,\quad G=0.0765,\quad M=7.5515,\quad Y=1.2945,\quad\sigma=0.
\end{align*}

\begin{figure}[hbt]
    {\par \centering
    \includegraphics[width=7.0cm,height=7.0cm]{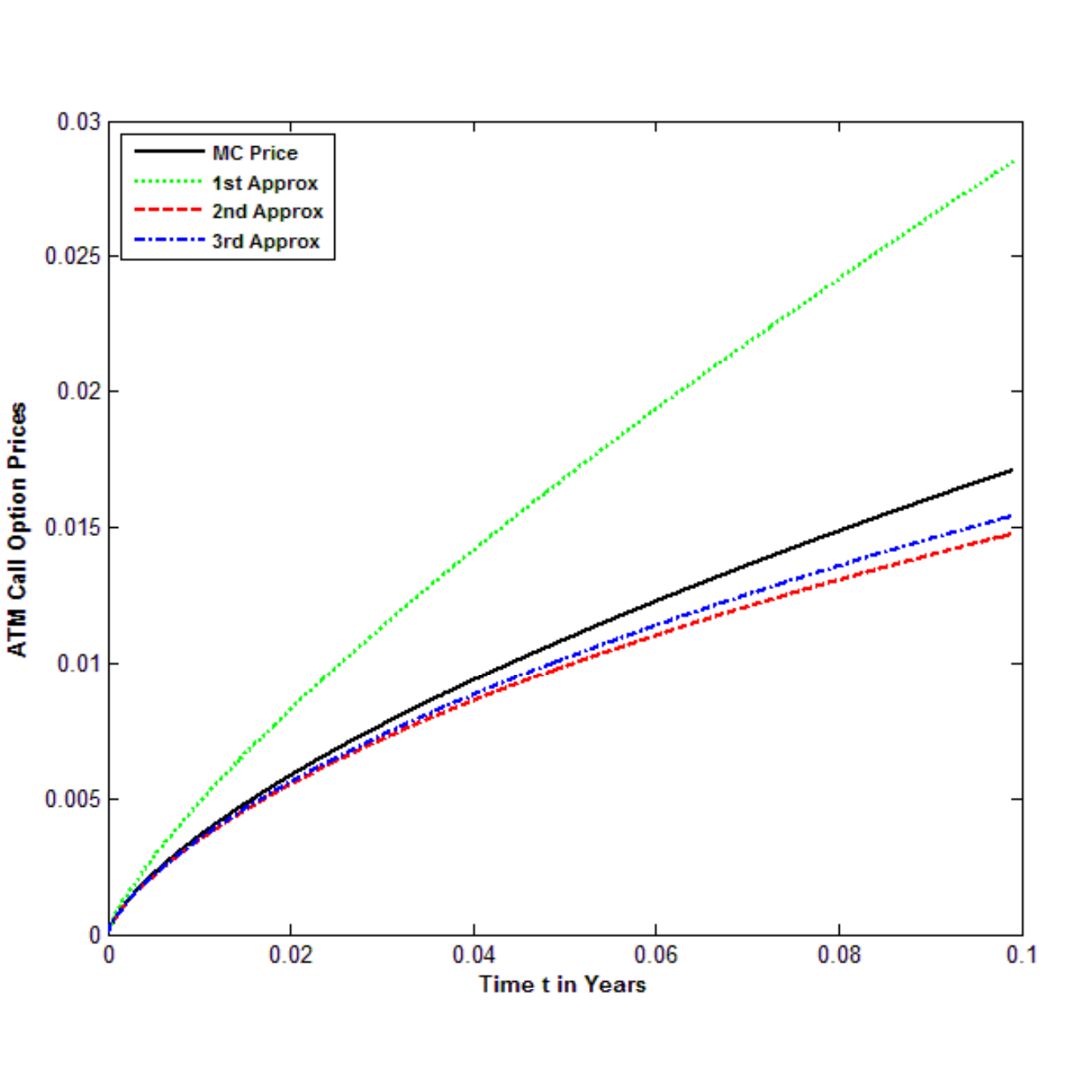}\hspace{1 cm}
    \includegraphics[width=7.0cm,height=7.0cm]{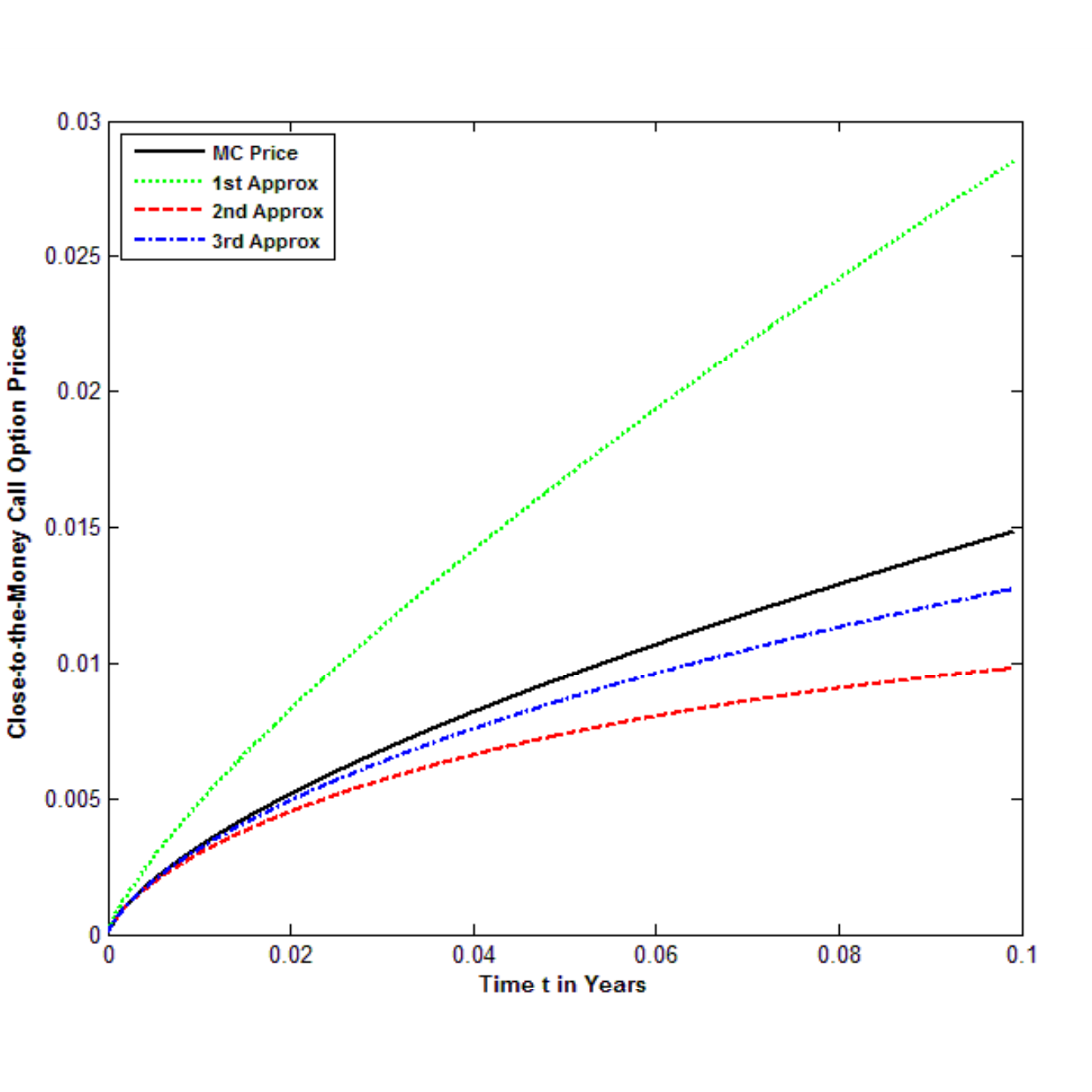}
    \par}\vspace{-0.5 cm}
    \caption{Comparisons of CGMY ATM and close-to-the-money call option prices with the first-, second- and third-order approximations. In both panels, $C=0.0244$, $G=0.0765$, $M=7.5515$, $Y=1.2945$, $\sigma=0$, {\Red and $\kappa_{t}=e_{1}t+e_{2}t^{2-1/Y}$}. In the left panel, $e_{1}=e_{2}=0$, while in the right panel, $e_{1}=0.1$ and $e_{2}=-0.1$.}\label{Fig:PureCGMYC00244}
\end{figure}

\begin{figure}[hbt]
    {\par \centering
    \includegraphics[width=7.0cm,height=7.0cm]{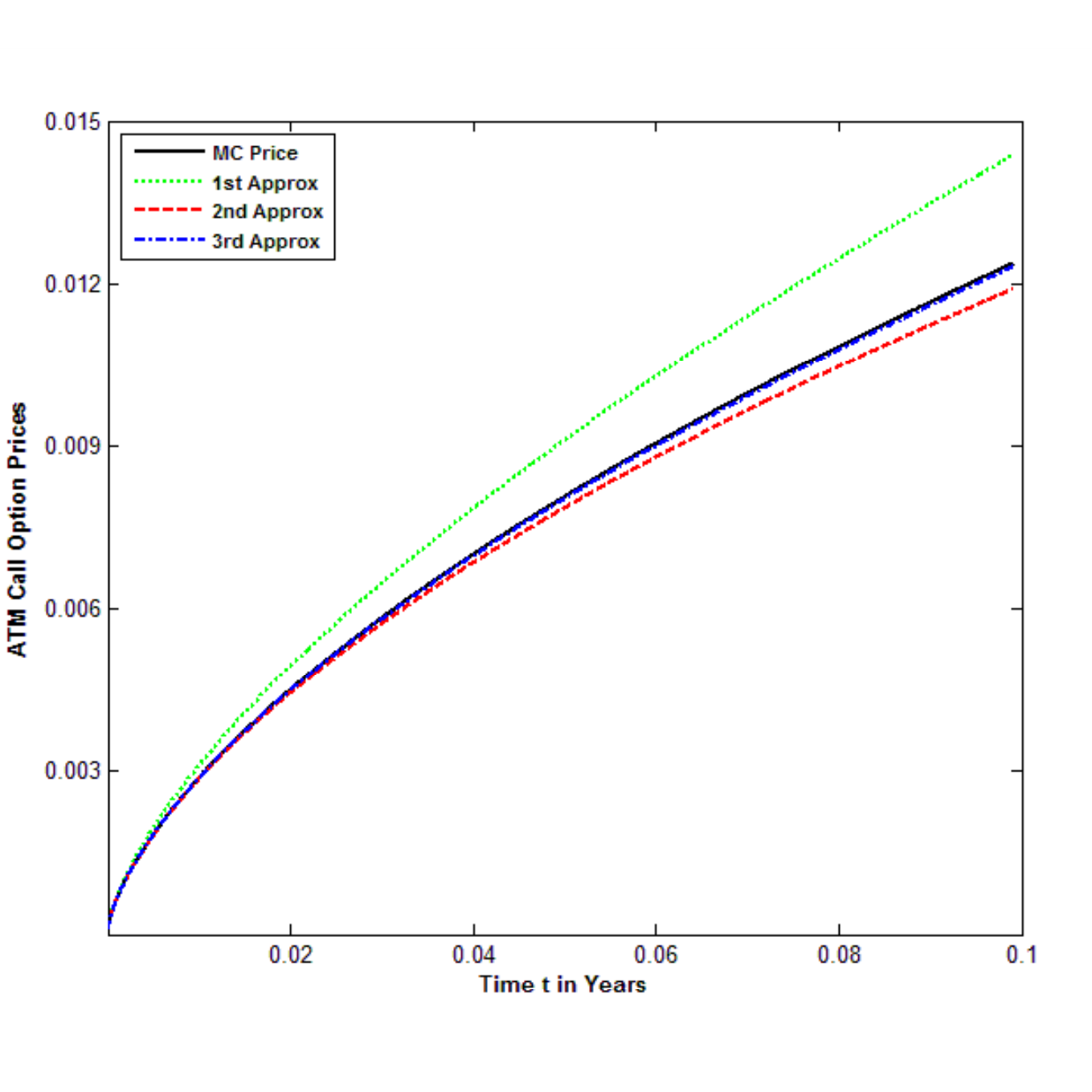}\hspace{1 cm}
    \includegraphics[width=7.0cm,height=7.0cm]{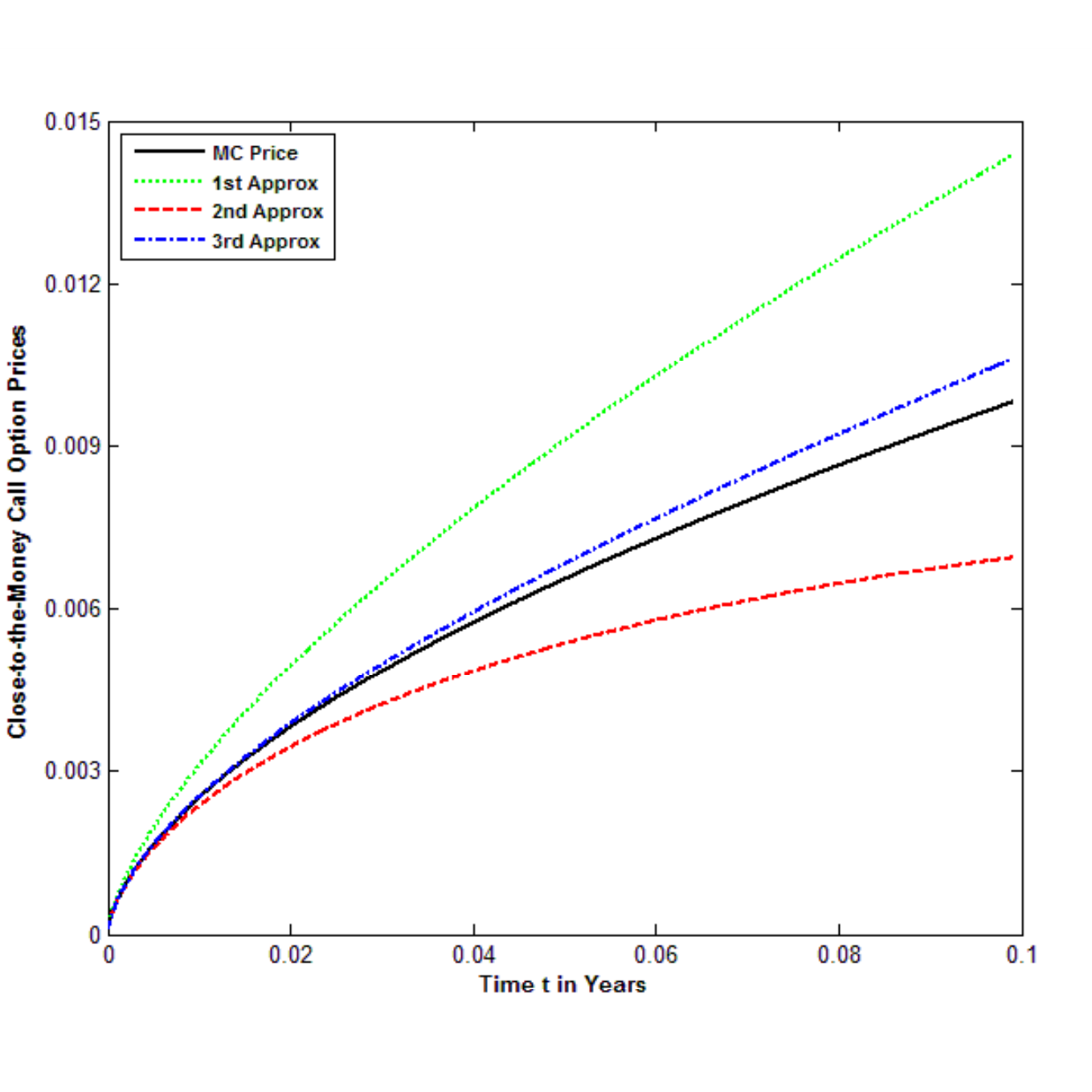}
    \par}\vspace{-0.5 cm}
    \caption{Comparisons of CGMY ATM and close-to-the-money call option prices with the first-, second- and third-order approximations. In both panels, $C=0.0066$, {$G=0.41$, $M=1.932$}, $Y=1.5$, $\sigma=0$, {\Red and $\kappa_{t}=e_{1}t+e_{2}t^{2-1/Y}$}. In the left panel, $e_{1}=e_{2}=0$, while in the right panel, $e_{1}=0.1$ and $e_{2}=-0.1$.}\label{Fig:PureCGMYC00066}
\end{figure}

\begin{figure}[hbt]
    {\par \centering
    \includegraphics[width=7.0cm,height=7.0cm]{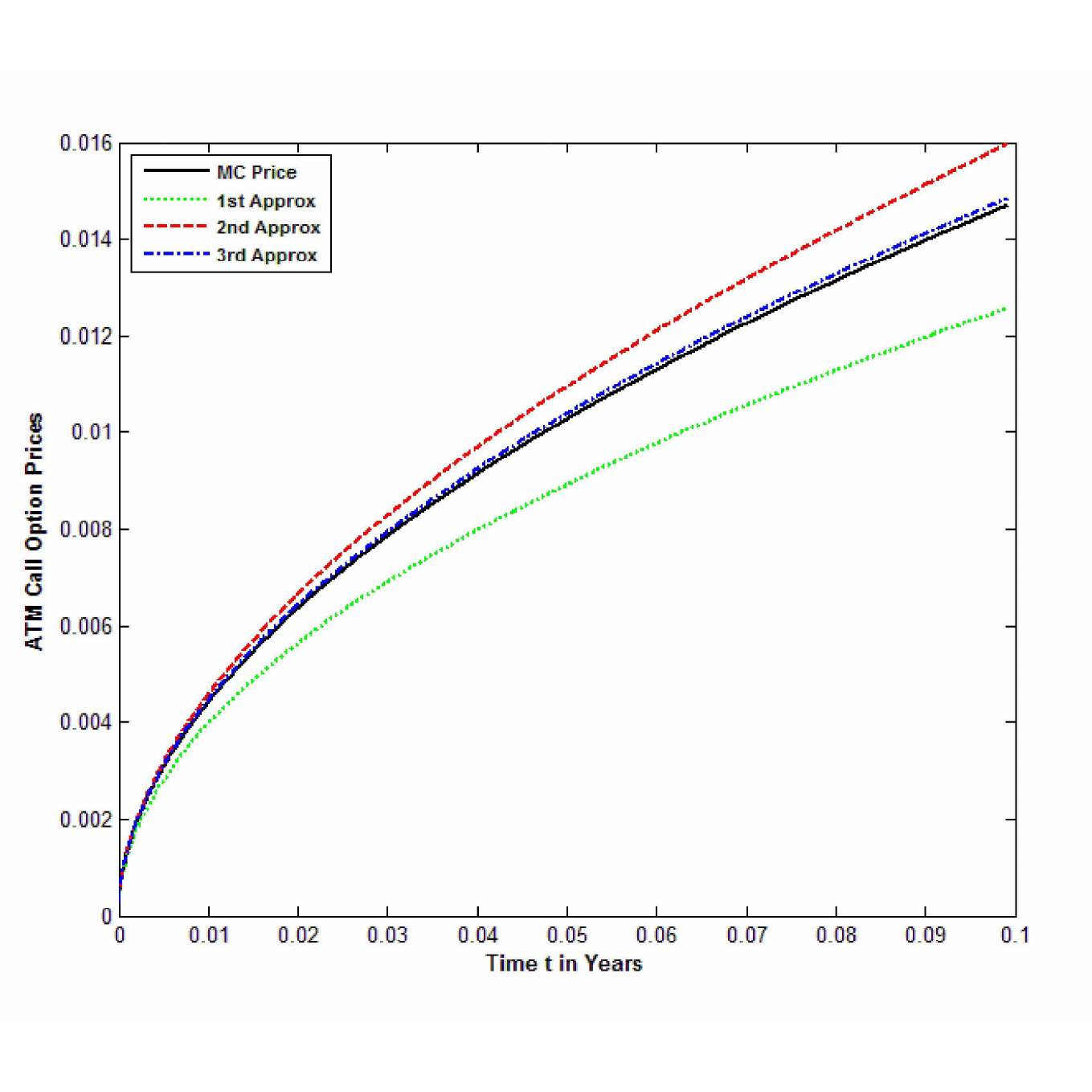}\hspace{1 cm}
    \includegraphics[width=7.0cm,height=7.0cm]{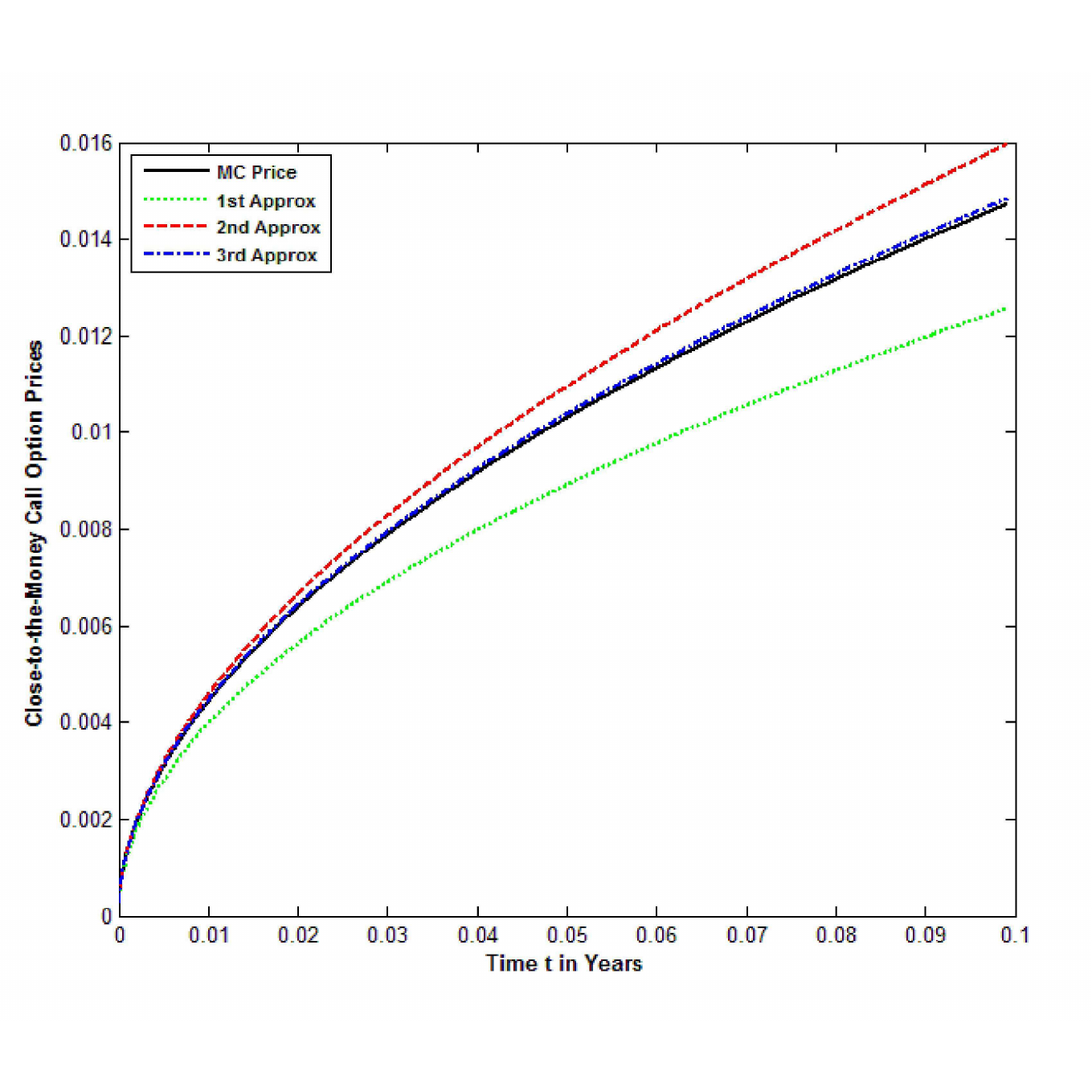}
    \par}\vspace{-0.5 cm}
    \caption{Comparisons of CGMY ATM and close-to-the-money call option prices with the first-, second- and third-order approximations. In both panels, $C=0.00265$, {$G=0.41$, $M=1.932$}, $Y=1.5$, $\sigma=0.1$, {\Red and $\kappa_{t}=e_{1}t+e_{2}t^{5/2-Y}$}. In the left panel, $e_{1}=e_{2}=0$, while in the right panel, $e_{1}=0.1$ and $e_{2}=-0.1$.}\label{Fig:GenCGMY}
\end{figure}

We use $100,000$ samples to simulate each of the MC-based option prices. The Figures \ref{Fig:PureCGMYC00244}-\ref{Fig:GenCGMY} compare the first-, second- and third-order approximations, as given in Theorem \ref{thm:ThirdOrderAsyPureCGMY} and Theorem \ref{thm:ThirdOrderAsyCGMYB}, to the prices based on the MC method introduced above, for both the pure-jump CGMY model and the mixed CGMY model. In all cases, the third-order approximation is much more accurate than the first- and the second-order approximations, for a time $t$ as large as one month.

Moreover, Table \ref{Tab:ElapsedTime} {summarizes} the elapsed time, in seconds, in simulating the MC-based prices as well as the first-, second- and third-order approximations in all cases. As expected, our asymptotic approximations are much more efficient than MC simulations, since all coefficients in our approximations are only made of simple algebraic computations of model parameters, except for $d_{32}$ in the pure-jump case.
\begin{center}
{\small
\begin{tabular}{|c|c|c|c|c|c|c|}
\hline
& & & & & & \\
Prices Models & Figure \ref{Fig:PureCGMYC00244} Left & Figure \ref{Fig:PureCGMYC00244} Right & Figure \ref{Fig:PureCGMYC00066} Left & Figure \ref{Fig:PureCGMYC00066} Right & Figure \ref{Fig:GenCGMY} Left & Figure \ref{Fig:GenCGMY} Right \\
& & & & & & \\
\hline
& & & & & & \\
MC-Based Prices & $183.70$ & $183.48$ & $182.83$ & $182.81$ & $150.33$ & $150.48$ \\
& & & & & & \\
\hline
& & & & & & \\
1st-order Approx & $3.67\times 10^{-4}$ & $7.21\times 10^{-4}$ & $3.61\times 10^{-4}$ & $4.41\times 10^{-4}$ & $1.24\times 10^{-4}$ & $1.46\times 10^{-4}$ \\
& & & & & & \\
\hline
& & & & & & \\
2nd-order Approx & $4.35\times 10^{-4}$ & $1.12\times 10^{-3}$ & $4.02\times 10^{-4}$ & $4.83\times 10^{-3}$ & $9.06\times 10^{-4}$ & $9.42\times 10^{-4}$ \\
& & & & & & \\
\hline
& & & & & & \\
3rd-order Approx & $8.19\times 10^{-4}$ & $2.11\times 10^{-3}$ & $0.97$ & $0.98$ & $1.90\times 10^{-3}$ & $1.94\times 10^{-3}$ \\
& & & & & & \\
\hline
\end{tabular}
}
\captionof{table}{Comparisons of the elapsed time (in seconds) of MC-based ATM and close-to-the-money call option prices with the first-, second- and third-order approximations under both the pure-jump CGMY model and the mixed CGMY model.}\label{Tab:ElapsedTime}
\end{center}

The computation of $d_{32}$, as given in (\ref{eq:3rdCoefPureCGMY2}), involves a double integral, which is numerically unstable if we first compute the tail probability as a function of $w$ using the MC method, and then evaluate the integral with respect to $w$. Instead, we will apply a two-dimensional MC method to evaluate the double integral term in $d_{32}$, which is denoted as $\tilde{d}_{32}$. More precisely, let
\begin{align*}
g(u,w)=w\left[{\bf 1}_{\{u\geq w\}}-\frac{CM^{Y}}{Yw^{Y}}-\frac{C(G+1)^{Y}}{Yw^{Y}}\right],\quad u\in\mathbb{R},\quad w\geq 0,
\end{align*}
and let $V$ be an absolutely continuous random variable, supported on $[0,\infty)$ and with density $f$ under $\widetilde{\mathbb{P}}$, which is independent of $U:=Z_{1}^{+}+\widetilde{U}_{1}$. Then,
\begin{align*}
\tilde{d}_{32}=\widetilde{\mathbb{E}}\left(\frac{g(U,V)}{f(V)}\right).
\end{align*}
The choice of the random variable $V$ will affect the efficiency and the stability of the MC method. Here we choose $V$ to be a standard half-normal random variable, and simulate $\tilde{d}_{32}$ using $1000^{2}$ pairs of samples $(U,V)$. As shown in Figure \ref{Fig:PureCGMYC00066} as well as in the fourth and fifth column of Table \ref{Tab:ElapsedTime}, the third-order approximations are almost identical to the MC-based prices, for $t$ as large as one month, while the corresponding elapsed time is negligible compared to that of the MC-based prices.

\subsection{Relevance in Calibration}

{As indicated in the introduction, {\Red a} main {\Red application} of short-time asymptotics is {\Red to} the calibration of the model's parameters. In practice this is complicated by {\Red the} fact that option prices generally exhibit errors and {\Red that} only certain maturities $t$ and log-moneyness values $k$ are available. {\Red Hence}, at any given date, we can expect the observed option prices $\Pi^{*}(t_{i},\kappa_{j})$ {\Red to} be given by
\begin{align*} \Pi^{*}(t_{i},\kappa_{j})=\Pi(t_{i},\kappa_{j})+\varepsilon_{ij}=S_{0}\,\mathbb{E}\left(\left(e^{X_{t_{i}}}-e^{\kappa_{j}}\right)^{+}\right)+\varepsilon_{ij},\quad i=1,\dots,I,\quad j=1,\dots,J,
\end{align*}
for some random errors $\varepsilon_{ij}$. The goal in this section is to show {\Red that}, in spite of the {\Red obstacles just mentioned}, the approximations herein can be applied to calibrate some of the model's parameters. For illustration {\Red purposes}, we focus on the volatility $\sigma$ and the parameter $C$. As we shall see, this is not only feasible but, {\Red moreover}, the higher-order approximations developed here are crucial for this {\Red endeavor,} and {\Red the first- and second-order approximations are not enough to give these results}.

Let us assume that, at a given date, we have at our hand ATM option prices $\Pi_{i}^{*}:=\Pi^{*}(t_{i},0)$ at maturities $t_{i}$ ($i=1,\dots,I$) and an estimate of the index $Y$, say $\hat{Y}$. The latter can be obtained {\Red from} high-frequency index or equity data as shown for instance in~\cite{AitSahaliaJacod:2009} (see also~\cite{FigueroaLopez:2012}). {\Red Then, the} basic idea to estimate $\sigma$ consists {\Red in} fitting {the linear models below to the data:}
\begin{align}\label{Ap1LM}
\Pi_{i}^{*}&:=d_{1}t_{i}^{1/2}+\varepsilon_{i},\qquad \text{({\Red First-Order})},\\
\label{Ap2LM} \Pi_{i}^{*}&:=d_{1}t_{i}^{1/2}+d_{2}t_{i}^{\frac{3-\hat{Y}}{2}}+\varepsilon_{i},\quad \text{({\Red Second-Order})},\\
\label{Ap3LM} \Pi_{i}^{*}&:=d_{1}t_{i}^{1/2}+d_{2}t_{i}^{\frac{3-\hat{Y}}{2}}+d_{31}t_{i}+d_{32}t_{i}^{\frac{5}{2}-\hat{Y}}+\varepsilon_{i},\quad\text{({\Red Higher-Order})}.
\end{align}
Let us denote the resulting {least-squares error estimates} of $d_{1}$ based on the three models above by $\hat{d}_{1}^{(1)},\hat{d}_{1}^{(2)},\hat{d}_{1}^{(3)}$. Since, theoretically, $d_{1}=\sigma/\sqrt{2\pi}$, the natural estimates for $\sigma$ are then given by
\begin{equation}\label{DfnEstSigma0}
\hat{\sigma}^{(\ell)}=\sqrt{2\pi}\,\hat{d}_{1}^{(\ell)},\quad\ell=1,2,3.
\end{equation}
We need to keep in mind that these estimates are based on short-time asymptotics for the option prices, which suggest to consider only ``small" $t_{i}$. However, these will reduce the sample size, {\Red making} the estimates more sensitive to errors.

Let us now show some numerical assessment of the estimates above. We first need to decide on some suitable maturities $t_{i}$. Based on the closing bid and ask prices for S\&P 500 index options on January 2nd, 2014, we find close-to-the money call option prices for the following maturities (in years):
\begin{align*}
\{t_{i}\}_{i=1,\ldots,15}\in\{0.021,0.043,0.060,0.079,0.140,0.217,0.241,0.293,0.467,0.491,0.717,0.744,0.967,1.043,1.467\}.
\end{align*}
The data was obtained from the website \texttt{HistoricalOptionData.com} and is shown {on Figure \ref{FgData}} \footnote{In addition to traditional S\&P 500 index options (SPX), the data includes SPXQ (quarterly) and SPXW (weekly) options. The latter class was first introduced in 2005, and, by the end of 2014, it accounted for over 40\% of the overall trading of S\&P 500 options on the CBOE.}. We then simulate ATM option prices at the above maturities using the parameter setting in (\ref{val2}) (borrowed from~\cite{AndersenLipton:2013}). The option prices and the approximations in (\ref{Ap1LM})-(\ref{Ap3LM}) with $\hat{Y}=1.5$ are shown {\Red on} the left panel of Figure \ref{FgATMP} (note that the high-order approximation almost overlaps with the option price). Now, since we {\Red do not} really have {\Red the exact value of $Y$ at our disposal}, we compute the estimates of $\hat{\sigma}^{(\ell)}$ for the range of values $\hat{Y}\in[1.1,1.8]$ to assess their sensitivity to the value $\hat{Y}$. The results are shown in Figure \ref{FgEstSigma1}. As shown therein, the estimate $\hat\sigma^{(3)}$ is relatively accurate for a large range of values of $\hat{Y}$ {(it ranges from $0.0968$ to $0.1053$)}, while the first- and second-order estimates are not. This fact would allow us to determine good estimates $\sigma$ even if our estimate $\hat{Y}$ is not very accurate.

We can also apply a similar idea to estimate the parameter $C$ based on the estimate of $d_{2}$, which theoretically {\Red is equal to}
\begin{align*}
d_{2}:=\frac{C2^{\frac{1-Y}{2}}\sigma^{1-Y}}{\sqrt{\pi}Y(Y-1)}\Gamma\left(1-\frac{Y}{2}\right)=:C\sigma^{1-Y}m_{Y}.
\end{align*}
Concretely, if $\hat{d}_{2}^{(2)}$ and $\hat{d}_{2}^{(3)}$ are the estimated values $d_{2}$ based on the regressions models (\ref{Ap2LM})-(\ref{Ap3LM}), then we set
\begin{equation}\label{EstCDfn}
{\Green \widehat{C}}^{(\ell)}=\hat{d}_{2}^{(\ell)}\left(\hat{\sigma}^{(\ell)}\right)^{\hat{Y}-1}m_{\hat{Y}}^{-1}, \quad {\Green \ell=2,3.}
\end{equation}
The resulting estimates applied to the simulated data of {\Red the} left panel in Figure \ref{FgATMP} are shown {\Red on} the left panel of Figure \ref{FgEstC1}. Unlike the estimate of $\sigma^{(3)}$, the estimate {\Green $\widehat{C}^{(3)}$ is more sensitive} to the estimate $\hat{Y}$, but when $\hat{Y}\approx 1.5$, the estimate seems accurate, {\Green while the estimate $\widehat{C}^{(2)}$ based on the second-order approximation (\ref{Ap1LM}) grossly underestimate $C$ even for $\hat{Y}=1.5$.}
	
To finish this section, we apply the estimators above to the market option data {shown on the table of Figure \ref{FgData}. For the first  {\Green {\Red six} maturities (1/10/14-3/22/14)}, the closest strike to the spot price {\Green $S_{0}=1831.98$ is $K=1830$}, while for the rest {\Red it} is 1825. For the first 4 maturities we take the closest-to-the-money option prices (those with $K=1830$), while for the rest we take the option prices corresponding to $K=1825$, because {\Green either} there is no option with strike $K=1830$ or because {\Green the option with $K=1825$ has larger volume than the one with $K=1830$, as in the case of options with maturities 2/22/14 and 3/22/14}. The mid prices of the selected data points against maturities are shown on the right panel of Figure \ref{FgATMP}. {\Green We also {\Red ran} the estimates below taking the options with strike $K=1825$ for all maturities  and {\Red essentially obtained} the same results.} The estimates of  $\hat\sigma^{(\ell)}$, $\ell=1,2,3$,  for different values of $\hat{Y}$ are shown on the right-panel of Figure \ref{FgEstSigma1}. As seeing therein, the volatility estimates $\hat{\sigma}^{(3)}$ are relatively stable for $\hat{Y}$ values in the range $[1.1,1.8]$ {({\Red they range} from $0.0898$ to $0.1121$)}. The estimates based on the first-order approximation (\ref{Ap1LM}) are much higher, while those based on the second-order approximation (\ref{Ap2LM}) are more sensitive to the {\Red estimates} of $\hat{Y}$. We also consider the estimates of $C$ defined in (\ref{EstCDfn}). The results are shown on the right-panel of Figure \ref{FgEstC1}. {\Green As with the simulated results}, the estimate ${\Green \widehat{C}}^{(3)}$ is more sensitive to the value $\hat{Y}$ {(it ranges from $0.0013$ to $0.0054$)}, and {\Red it} is expected that {\Red an accurate estimate of $\hat{Y}$ would be able to accurately estimate $C$}. The estimate ${\Green \widehat{C}}^{(2)}$ {\Red based only} on the {\Red second-order} approximation is more stable but, from our simulations, it is expected to {\Red sharply underestimate} $C$. For completeness, we also {\Red show} the analogous estimators based on the expansion (\ref{eq:3rdExpPureCGMY1}) {\Red using only} the $d_{1}$, $d_{2}$, and $d_{31}$ terms, and the estimator {\Red using only} the $d_{1}$, $d_{2}$, and $d_{32}$ terms.}

\begin{figure}[hp]
\vspace{-1 cm}
    {\par \centering
    \includegraphics[width=15.0cm,height=17.0cm]{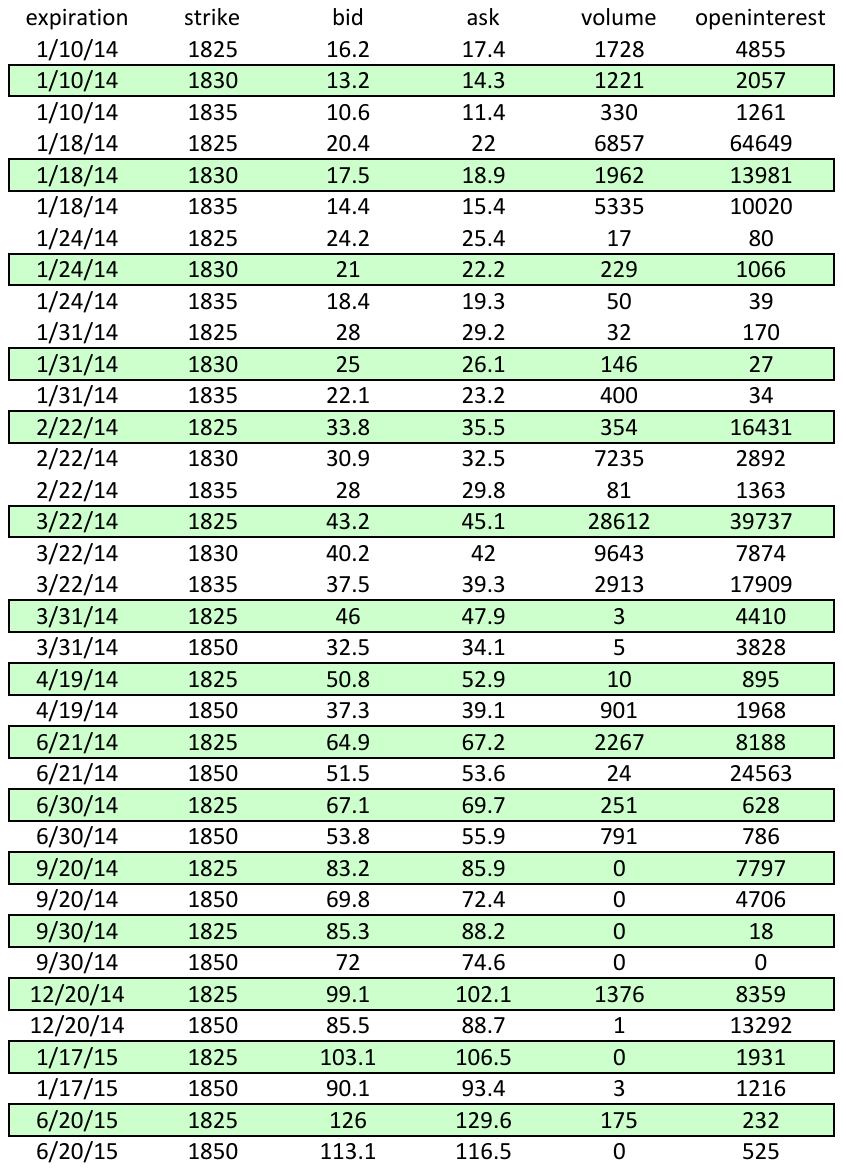}
    \par}\vspace{-3 cm}
    \caption{Close-to-the-money SPX Call Option Data on Jan/2/2014 when the underlying was at 1831.98. The emphasized rows (in green) consists of those used in our calibration.}\label{FgData}
\end{figure}

\begin{figure}[hbt]
    {\par \centering
    \includegraphics[width=8.0cm,height=7.0cm]{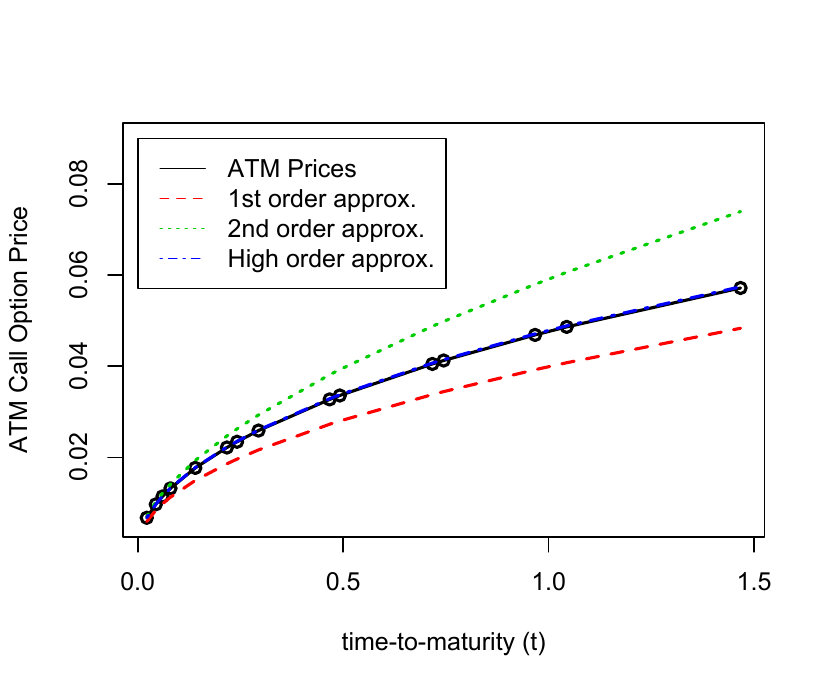}
\hspace{1 cm}        \includegraphics[width=8.0cm,height=7.0cm]{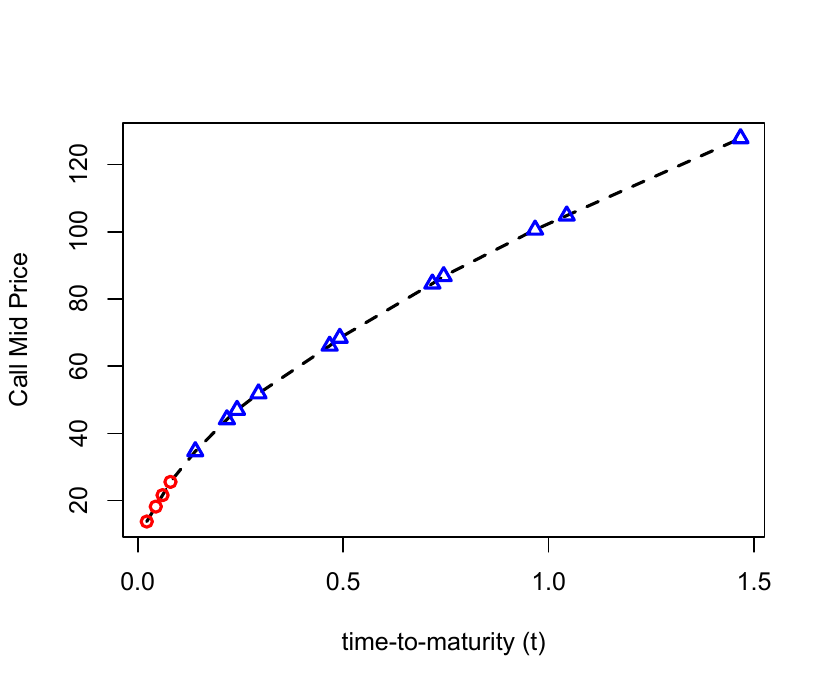}
    \par}\vspace{-0.5 cm}
    \caption{Right Panel: Comparisons of CGMY ATM call option prices to its first-, second- and third-order approximations ($C=0.00265$, $G=0.41$, $M=1.93$, $Y=1.5$, $\sigma=0.1$). Left Panel: Close-to-the-money SPX call prices on Jan. 2, 2014 for different time-to-maturities: $S_{0}=1831.98$ and $K=1830$ (red circles) or $K=1825$ (blue triangles).}\label{FgATMP}
\end{figure}
\begin{figure}[hbt]
    {\par \centering
    \includegraphics[width=8.0cm,height=7.0cm]{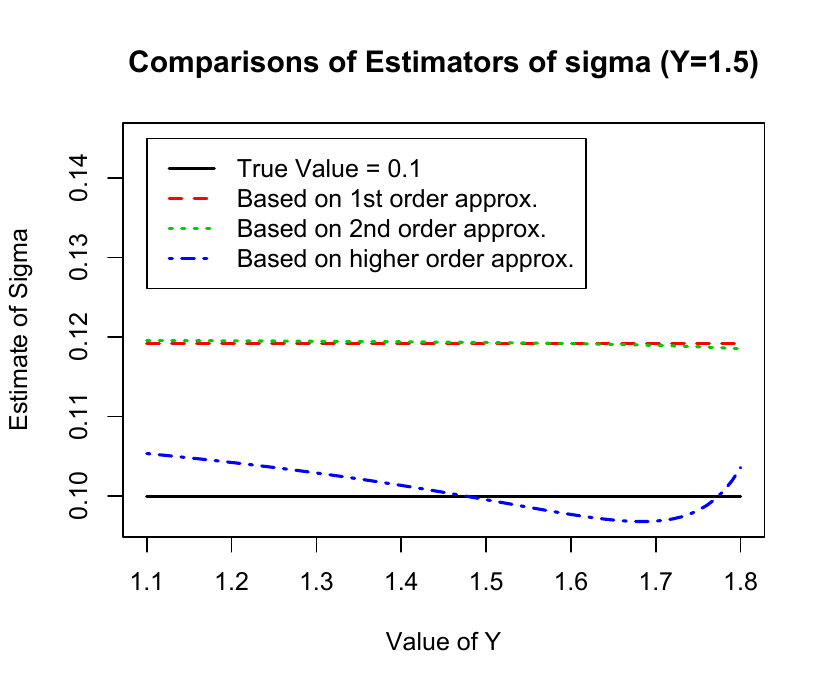}\hspace{1 cm}
    \includegraphics[width=8.0cm,height=7.0cm]{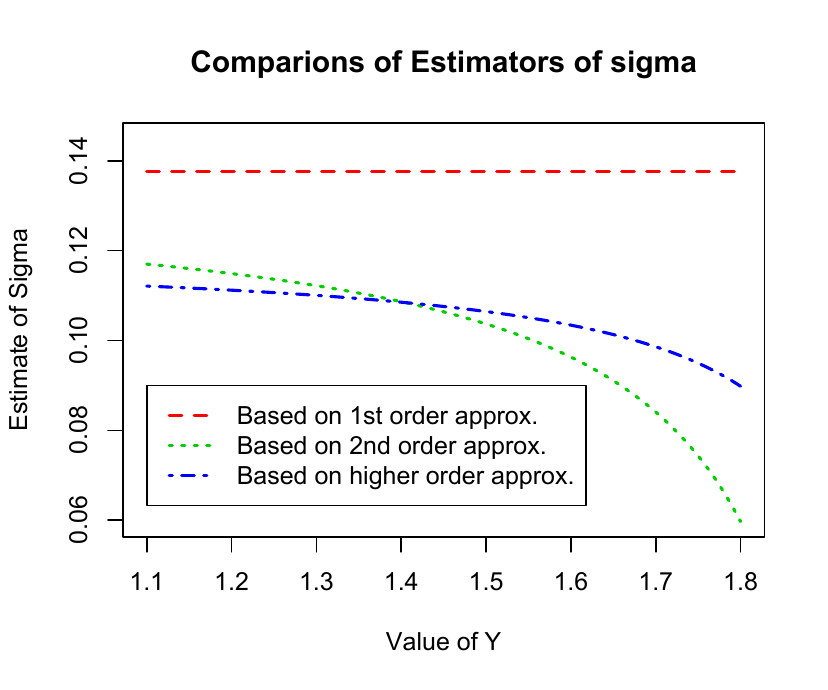}
    \par}\vspace{-0.5 cm}
    \caption{Comparisons of estimates $\hat{\sigma}^{(\ell)}$ of $\sigma$ as defined in (\ref{DfnEstSigma0}). Left Panel: Estimates based on simulated CGMY ATM Option prices with parameters $C=0.00265$, $G=0.41$, $M=1.93$, $Y=1.5$, $\sigma=0.1$. Right panel: Estimates based on the close-to-the money SPX call option prices plotted in the right panel of Figure \ref{FgATMP}.}\label{FgEstSigma1}
\end{figure}
\begin{figure}[hbt]
    {\par \centering
    \includegraphics[width=8.0cm,height=7.0cm]{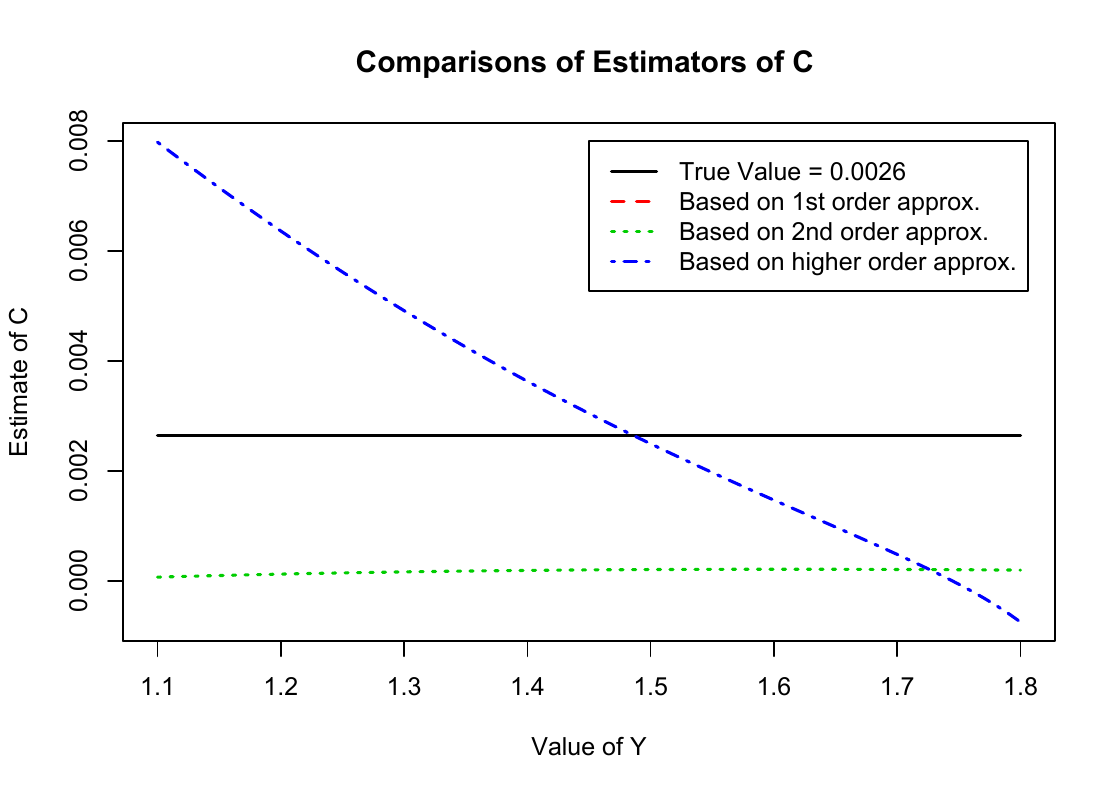}\hspace{1 cm}
    \includegraphics[width=8.0cm,height=7.0cm]{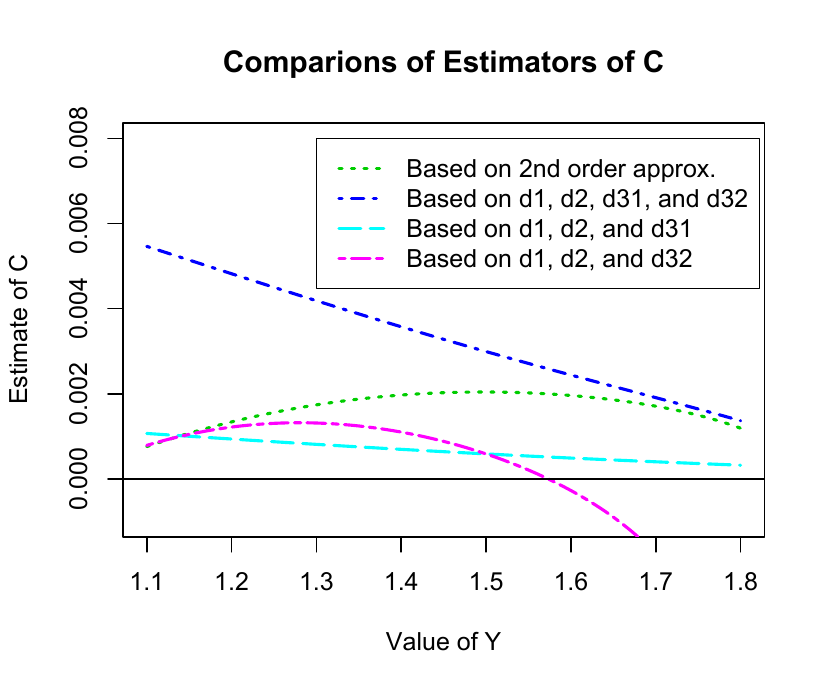}
    \par}\vspace{-0.5 cm}
    \caption{Comparisons of estimates ${\Green \widehat{C}}^{(\ell)}$ of $C$ as defined in (\ref{EstCDfn}). Left Panel: Estimates based on simulated CGMY ATM Option prices with parameters $C=0.00265$, $G=0.41$, $M=1.93$, $Y=1.5$, $\sigma=0.1$. Right panel: Estimates based on the close-to-the money SPX call option prices plotted in the right panel of Figure \ref{FgATMP}.}\label{FgEstC1}
\end{figure}

\appendix

\section{Proofs of the Main Results}\label{proofA}

For notational simplicity, throughout all the proofs, we fix $S_{0}=1$.

\medskip
\noindent
\textbf{Proof of Lemma \ref{thm:ThirdOrderAsyCGMYBLemma}.}
We show the proof for $R_{t}^{(0)}$ (the proof for $R_{t}^{(1)}$ is similar). To start,
\begin{align*}
R_{t}^{(0)}=\frac{1}{2}\int_{0}^{1}\int_{\mathbb{R}}\frac{t}{\sqrt{2\pi\sigma^{2}}}e^{-\frac{t^{2}z^{2}u^{2}}{2\sigma^{2}}}|z|^{2}\left(p_{Z}(z)-C|z|^{-Y-1}\right)dz\,du.
\end{align*}
Next, denoting the characteristic function of $Z_{1}$ by $\hat{p}_{Z}(x)$, note that
\begin{align*}
p_{Z}(z)=\mathcal{F}\left(\frac{1}{\sqrt{2\pi}}\,\hat{p}_{Z}\right)(z),\quad z^{2}p_{Z}(z)=\mathcal{F}\left(\frac{-1}{\sqrt{2\pi}}\,\hat{p}_{Z}''\right)(z),
\end{align*}
where $\mathcal{F}(h)(z):=\int_{\mathbb{R}}e^{-ivz}h(v)dv/\sqrt{2\pi}$ denotes the Fourier transformation of $h\in L_{1}(\mathbb{R})$. Also, regarding $|x|^{Y-2}$ as a tempered distribution, it is known that $|z|^{1-Y}=\mathcal{F}(K^{-1}|x|^{Y-2})(z)$,
with $K:=-2\sin(\pi(Y-2)/2)\Gamma(Y-1)/\sqrt{2\pi}$. In particular,
\begin{align*}
R_{t}^{(0)}&=\frac{1}{2}\int_{0}^{1}\int_{\mathbb{R}}\mathcal{F}\left(\frac{t}{\sqrt{2\pi\sigma^{2}}}e^{-\frac{t^{2}z^{2}u^{2}}{2\sigma^{2}}}\right)(x)\left(-\frac{1}{\sqrt{2\pi}}\,\hat{p}''_{Z}(x)-\frac{C}{K}|x|^{Y-2}\right)dx\,du\\
&=-\frac{1}{2\sqrt{2\pi}}\int_{0}^{1}\int_{\mathbb{R}}u^{-1}e^{-\frac{\sigma^{2}x^{2}}{2t^{2}u^{2}}}\left(\frac{1}{\sqrt{2\pi}}\,\hat{p}''_{Z}(x)+\frac{C}{K}|x|^{Y-2}\right)dx\,du.
\end{align*}
Since $\hat{p}_{Z}(x)=e^{-c|x|^{Y}}$ with $c:=2C|\cos(\pi Y/2)|\Gamma(-Y)$ and $C/K=cY(Y-1)/\sqrt{2\pi}$, differentiation gives
\begin{align}\label{eq:SecondDecomB133}
R_{t}^{(0)}=\frac{-c^{2}Y^{2}}{2\pi}\!\int_{0}^{1}\!\int_{0}^{\infty}e^{-\frac{\sigma^{2}x^{2}}{2t^{2}u^{2}}}\frac{x^{2Y-2}}{u}e^{-cx^{Y}}dx\,du+\frac{-cY(Y-1)}{2\pi}\int_{0}^{1}\!\int_{0}^{\infty}e^{-\frac{\sigma^{2}x^{2}}{2t^{2}u^{2}}}\frac{x^{Y-2}}{u}\left(1-e^{-cx^{Y}}\right)dx\,du.
\end{align}
For the first term in (\ref{eq:SecondDecomB133}), which we denote by $R_{t}^{(01)}$, we change variables from $x$ to $v=\sigma x/tu$ to get
\begin{align*}
R_{t}^{(01)}=-\frac{c^{2}Y^{2}}{2\pi\sigma^{2Y-1}}t^{2Y-1}\int_{0}^{1}\left(\int_{0}^{\infty}e^{-\frac{v^{2}}{2}}v^{2Y-2}\exp\left(-\frac{c(tuv)^{Y}}{\sigma^{Y}}\right)dv\right)u^{2Y-2}\,du.
\end{align*}
Hence, by the dominated convergence theorem,
\begin{align}\label{eq:AsyB1331}
\lim_{t\rightarrow 0}t^{1-2Y}R_{t}^{(01)}(t)=-\frac{c^{2}Y^{2}}{2\sqrt{2\pi}(2Y-1)\sigma^{2Y-1}}\mathbb{E}\left(\left|W_{1}\right|^{2Y-2}\right)=-\frac{2C^{2}Y^{2}\cos^{2}\left(\frac{\pi Y}{2}\right)\Gamma^{2}(-Y)}{\sqrt{2\pi}(2Y-1)\sigma^{2Y-1}}{\mathbb{E}}\left(\left|W_{1}\right|^{2Y-2}\right).
\end{align}
Similarly, the asymptotic behavior of the second term in (\ref{eq:SecondDecomB133}), which we denote by $R_{t}^{(02)}$, is given by
\begin{align}\label{eq:AsyB1332}
\lim_{t\rightarrow 0}t^{1-2Y}R_{t}^{(02)}(t)\!=\!\frac{-c^{2}Y(Y-1)}{2\sqrt{2\pi}(2Y\!-\!1)\sigma^{2Y-1}}{\mathbb{E}}\left(\left|W_{1}\right|^{2Y-2}\right)=-\frac{2C^{2}Y(Y\!-\!1)\cos^{2}\!\left(\frac{\pi Y}{2}\right)\Gamma^{2}(-Y)}{\sqrt{2\pi}(2Y-1)\sigma^{2Y-1}}{\mathbb{E}}\left(\left|W_{1}\right|^{2Y-2}\right).
\end{align}
Combining (\ref{eq:AsyB1331}) and (\ref{eq:AsyB1332}), we get that $R^{(0)}_{t}\sim t^{2Y+k-1}E^{(0)}$, with $E^{(0)}$ given in (\ref{DfnE01}).\hfill $\Box$

\medskip
\noindent
\textbf{Proof of Theorem \ref{thm:ThirdOrderAsyCGMYB}.}
{Let $\widetilde{X}_{t}:=X_{t}-\kappa_{t}$. {\Red Then,}
\begin{align}\label{eq:DecomOPT}
\mathbb{E}\left(\left(S_{t}-e^{\kappa_{t}}\right)^{+}\right)=\mathbb{E}^{*}\left(1-e^{-\widetilde{X}_{t}^{+}}\right)=e^{\kappa_{t}}\int_{\kappa_{t}}^{\infty}e^{-v}\,\mathbb{P}^{*}\!\left(X_{t}\geq v\right)dv=:e^{\kappa_{t}}(G_{2}(t)-G_{1}(t)),
\end{align}
where
\begin{align}\label{eq:DefG1G2}
G_{1}(t):=\int_{0}^{\kappa_{t}}e^{-v}\,\mathbb{P}^{*}\!\left(X_{t}\geq v\right)dv,\quad G_{2}(t):=\int_{0}^{\infty}e^{-v}\,\mathbb{P}^{*}\!\left(X_{t}\geq v\right)dv=\sqrt{t}\int_{0}^{\infty}e^{-\sqrt{t}w}\,\mathbb{P}^{*}\!\left(X_{t}\geq t^{-1/2}w\right)dw.
\end{align}
For $G_{1}(t)$, note that, if $\kappa_{t}\neq{}0$, changing {\Red variables} to $w=v/\kappa_{t}$,
\begin{align*}
G_{1}(t)=\kappa_{t}\int_{0}^{1}e^{-\kappa_{t} w}\mathbb{P}^{*}\!\left(t^{-1/2}X_{t}\geq t^{-1/2}\kappa_{t}w\right)dv=\kappa_{t}\int_{0}^{1}\mathbb{P}^{*}\!\left(\sigma W_{1}\geq t^{-1/2}\kappa_{t}w\right)dw+o(\kappa_{t}),\quad {\Red t\rightarrow 0,}
\end{align*}
where {\Red we used the well-know fact that $t^{-1/2}X_{t}\stackrel{\mathcal{D}}{\to}\sigma W_{1}$, as $t\to{}0$, and the fact that pointwise convergence of a sequence of distribution functions to a continuous distribution function implies uniform convergence. Indeed, under $\mathbb{P}^{*}$,
$t^{-1/2}X_{t}=\sigma W^{*}_{1}+t^{(2-Y)/2Y}t^{-1/Y}L^{*}_{t}$ and $t^{-1/Y}L^{*}_{t}$ converges in distribution to a symmetric strictly $Y$-stable random variable under $\mathbb{P}^{*}$ (cf.~\cite[Theorem 3.1]{Rosinski:2007}).} Now, {\Red to} handle $G_{2}(t)$, {\Red fix} $\tilde{\gamma}_{t}:=t^{1/2}\tilde{\gamma}$ and consider
\begin{align*}
\Delta_{0}(t)&:=\frac{1}{\sqrt{t}}\,G_{2}(t)-d_{1}=\int_{-\tilde{\gamma}_{t}}^{\infty}e^{-\sqrt{t}y-\sqrt{t}\tilde{\gamma}_{t}}\mathbb{P}^{*}\left(\sigma W_{1}^{*}\geq y-t^{-\frac{1}{2}}Z_{t}\right)dy-\int_{0}^{\infty}\mathbb{P}^{*}(\sigma W_{1}^{*}\geq y)\,dy.
\end{align*}
By changing the probability measure $\mathbb{P}^{*}$ to $\widetilde{\mathbb{P}}$   and using (\ref{DcmLL}) as well as the self-similarity of $((Z_{t},U_{t}))_{t\geq 0}$,
\begin{align}
\Delta_{0}(t)&=e^{-(\eta t+\sqrt{t}\tilde{\gamma}_{t})}\int_{0}^{\infty}e^{-\sqrt{t}y}\left(\widetilde{\mathbb{E}}\left(e^{-t^{\frac{1}{Y}}\widetilde{U}_{1}}{\bf 1}_{\left\{\sigma W_{1}^{*}\geq y-t^{\frac{1}{Y}-\frac{1}{2}}Z_{1}\right\}}\right)-\widetilde{\mathbb{E}}\left(e^{-t^{\frac{1}{Y}}\widetilde{U}_{1}}{\bf 1}_{\left\{\sigma W_{1}^{*}\geq y\right\}}\right)\right)dy\nonumber\\
&\quad+\int_{0}^{\infty}\left(e^{-\sqrt{t}\tilde{\gamma}_{t}-\sqrt{t}y}-1\right)\widetilde{\mathbb{P}}(\sigma W_{1}^{*}\geq y)\,dy+e^{-\sqrt{t}\tilde{\gamma}_{t}}\int_{-\tilde{\gamma}_{t}}^{0}e^{-\sqrt{t}y}\,\widetilde{\mathbb{E}}\left(e^{-t^{\frac{1}{Y}}\widetilde{U}_{1}-\eta t}{\bf 1}_{\left\{\sigma W_{1}^{*}\geq y-t^{\frac{1}{Y}-\frac{1}{2}}Z_{1}\right\}}\right)dy\nonumber\\
\label{DecomBt} &=:A_{1}(t)+A_{2}(t)+A_{3}(t).
\end{align}
It is not hard to see that
\begin{align*}
A_{2}(t)\sim -\frac{\sigma^{2}}{4}t^{1/2},\quad A_{3}(t)=\frac{\tilde{\gamma}}{2}t^{\frac{1}{2}}+o\left(t^{\frac{1}{2}}\right),\quad t\rightarrow 0.
\end{align*}
To} study the asymptotic behavior of $A_{1}(t)$, we first decompose it as:
\begin{align}
A_{1}(t)&=e^{-(\eta t+\sqrt{t}\tilde{\gamma}_{t})}\widetilde{\mathbb{E}}\left(e^{-t^{\frac{1}{Y}}\widetilde{U}_{1}}{\bf 1}_{\left\{W_{1}^{*}\geq 0,\,\sigma W_{1}^{*}+t^{\frac{1}{Y}-\frac{1}{2}}Z_{1}\geq 0\right\}}\int_{\sigma W_{1}^{*}}^{\sigma W_{1}^{*}+t^{\frac{1}{Y}-\frac{1}{2}}Z_{1}}e^{-\sqrt{t}y}\,dy\right)\nonumber\\
&\quad-e^{-(\eta t+\sqrt{t}\tilde{\gamma}_{t})}\widetilde{\mathbb{E}}\left(e^{-t^{\frac{1}{Y}}\widetilde{U}_{1}}{\bf 1}_{\left\{0\leq\sigma W_{1}^{*}\leq-t^{\frac{1}{Y}-\frac{1}{2}}Z_{1}\right\}}\int_{0}^{\sigma W_{1}^{*}}e^{-\sqrt{t}y}\,dy\right)\nonumber\\
&\quad+e^{-(\eta t+\sqrt{t}\tilde{\gamma}_{t})}\widetilde{\mathbb{E}}\left(e^{-t^{\frac{1}{Y}}\widetilde{U}_{1}}{\bf 1}_{\left\{0\leq-\sigma W_{1}^{*}\leq t^{\frac{1}{Y}-\frac{1}{2}}Z_{1}\right\}}\int_{0}^{\sigma W_{1}^{*}+t^{\frac{1}{Y}-\frac{1}{2}}Z_{1}}e^{-\sqrt{t}y}\,dy\right)\nonumber\\
\label{DecomB1} &=:e^{-(\eta t+\sqrt{t}\tilde{\gamma}_{t})}\left(B_{1}(t)-B_{2}(t)+B_{3}(t)\right).
\end{align}
Each of the above terms is now analyzed individually in three subsequent steps.

\smallskip
\noindent
\textbf{Step 1.} First, by the change of variable $u=t^{1/2-1/Y}y-\sigma t^{1/2-1/Y}W_{1}^{*}+\widetilde{U}_{1}$, Fubini's theorem, the independence of $W^{*}_{1}$ and $(Z_{1},\widetilde{U}_{1})$, and the symmetry of $Z_{1}$, we have
\begin{align}
B_{1}(t)&=t^{\frac{1}{Y}-\frac{1}{2}}\widetilde{\mathbb{E}}\left(e^{-\sqrt{t}\sigma W_{1}^{*}}{\bf 1}_{\left\{W_{1}^{*}\geq 0\right\}}\right)\int_{\mathbb{R}}\left(e^{-t^{\frac{1}{Y}}u}-1\right)\widetilde{\mathbb{P}}\left(Z_{1}\geq 0,\,\widetilde{U}_{1}\leq u\leq\widetilde{U}_{1}+Z_{1}\right)du\nonumber\\
&\quad-t^{\frac{1}{Y}-\frac{1}{2}}\int_{0}^{\infty}\left(\int_{\mathbb{R}}\left(e^{-t^{\frac{1}{Y}}u}-1\right)\widetilde{\mathbb{P}}\left(-t^{\frac{1}{2}-\frac{1}{Y}}w\leq Z_{1}\leq 0,\,\widetilde{U}_{1}+Z_{1}\leq u\leq\widetilde{U}_{1}\right)du\right)e^{-\sqrt{t}w}\frac{e^{-\frac{w^{2}}{2\sigma^{2}}}}{\sqrt{2\pi\sigma^{2}}}\,dw\nonumber\\
&\quad+t^{\frac{1}{Y}-\frac{1}{2}}\widetilde{\mathbb{E}}\left(Z_{1}{\bf 1}_{\left\{W_{1}^{*}\geq 0,\,Z_{1}\geq t^{\frac{1}{2}-\frac{1}{Y}}\sigma W_{1}^{*}\right\}}e^{-\sqrt{t}\sigma W_{1}^{*}}\right)\nonumber\\
\label{Eq:DfnC123} &=:B_{11}(t)-B_{12}(t)+B_{13}(t).
\end{align}
To analyze $B_{11}(t)$, we use arguments similar to those used to obtain (\ref{eq:AsyA3}) and (\ref{NdTp1}) to get
\begin{align*}
\int_{-\infty}^{0}\left(e^{-t^{\frac{1}{Y}}u}-1\right)\widetilde{\mathbb{P}}\left(Z_{1}\geq 0,\widetilde{U}_{1}\leq u\leq\widetilde{U}_{1}+Z_{1}\right)du&=t^{\frac{1}{Y}}\int_{-\infty}^{0}(-u)\,\widetilde{\mathbb{P}}\left(Z_{1}\geq 0,\,\widetilde{U}_{1}\leq u\leq\widetilde{U}_{1}+Z_{1}\right)du+o(t^{\frac{1}{Y}}),\\
\int_{0}^{\infty}\left(e^{-t^{\frac{1}{Y}}u}-1\right)\widetilde{\mathbb{P}}\left(Z_{1}\geq 0,\widetilde{U}_{1}\leq u\leq\widetilde{U}_{1}+Z_{1}\right)du&=-t^{1-\frac{1}{Y}}C\Gamma(-Y)\left[M^{Y}-\left(M^{*}\right)^{Y}\right]+O(t^{\frac{1}{Y}}),
\end{align*}
which imply that
\begin{align}\label{AsymDescC1}
B_{11}(t)=-\frac{1}{2}t^{\frac{1}{2}}C\Gamma(-Y)\left[M^{Y}-\left(M^{*}\right)^{Y}\right]+O\left(t^{\frac{2}{Y}-\frac{1}{2}}\right),\quad t\rightarrow 0.
\end{align}
It turns out that (see Appendix \ref{proofB}),
\begin{align}\label{AsympDescC2}
B_{12}(t)=o(t^{\frac{1}{2}}),\quad t\rightarrow 0.
\end{align}
Finally, we deal with $B_{13}(t)$, for which we study the asymptotic behavior of
\begin{align}\label{DfnDt}
\widetilde{B}_{13}(t)&:=t^{\frac{Y}{2}-1}B_{13}(t)-\frac{C\sigma^{1-Y}}{2(Y-1)}\widetilde{\mathbb{E}}\left(\left|W_{1}^{*}\right|^{1-Y}\right)\\
&\,=t^{\frac{Y}{2}+\frac{1}{Y}-\frac{3}{2}}\widetilde{\mathbb{E}}\left(\left(e^{-\sqrt{t}\sigma W_{1}^{*}}-1\right){\bf 1}_{\{W_{1}^{*}\geq 0\}}\int_{t^{\frac{1}{2}-\frac{1}{Y}}\sigma W_{1}^{*}}^{\infty}z\,p_{Z}(1,z)\,dz\right)\nonumber\\
\label{DecomJ15} &\quad\,+t^{\frac{Y}{2}+\frac{1}{Y}-\frac{3}{2}}\widetilde{\mathbb{E}}\left({\bf 1}_{\{W_{1}^{*}\geq 0\}}\int_{t^{\frac{1}{2}-\frac{1}{Y}}\sigma W_{1}^{*}}^{\infty}z\left(p_{Z}(1,z)-Cz^{-Y-1}\right)dz\right).
\end{align}
We claim that the first term in (\ref{DecomJ15}) is of order $O(\sqrt{t})$. Indeed, by (\ref{Asydenpz}), there exists $H_{1}>0$ such that, for any $z\geq H_{1}$, $p_{Z}(1,z)\leq 2Cz^{-Y-1}$. Hence, for any $w>0$,
\begin{align*}
\int_{t^{\frac{1}{2}-\frac{1}{Y}}w}^{\infty}\!zp_{Z}(1,z)dz\leq\int_{t^{\frac{1}{2}-\frac{1}{Y}}w}^{\infty}\!2Cu^{-Y}du+{\bf 1}_{\{t^{\frac{1}{2}-\frac{1}{Y}}w<H_{1}\}}H_{1}\widetilde{\mathbb{P}}\left(Z_{1}\geq t^{\frac{1}{2}-\frac{1}{Y}}w\right)\leq
t^{{\frac{3}{2}}-\frac{Y}{2}-\frac{1}{Y}}\left(\frac{2Cw^{1-Y}}{Y-1}+H_{1}^{Y}w^{1-Y}\right),
\end{align*}
where in the last inequality we used that $\widetilde{\mathbb{P}}(Z_{1}\geq {\Red t^{1/2-1/Y}}w)\leq (H_{1}/{\Red t^{1/2-1/Y}}w)^{Y-1}$, when ${\Red t^{1/2-1/Y}}w<H_{1}$. Now, the second term in (\ref{DecomJ15}) is nothing else than ${\Red t^{Y/2+1/Y-3/2}}R_{t^{(2-Y)/2Y}}^{(0)}$ and, thus, we can apply Lemma \ref{thm:ThirdOrderAsyCGMYBLemma} to get
\begin{align}\label{1stAsyJ15}
\lim_{t\rightarrow 0}t^{\frac{Y}{2}-1}\widetilde{B}_{13}(t)=-\frac{2C^{2}Y\cos^{2}\left(\frac{\pi Y}{2}\right)\Gamma^{2}(-Y)}{\sqrt{2\pi}\sigma^{2Y-1}}\widetilde{\mathbb{E}}\left(\left|W_{1}^{*}\right|^{2Y-2}\right)=:d_{31}'.
\end{align}
Combining (\ref{Eq:DfnC123}), (\ref{AsymDescC1})-(\ref{DfnDt}), and (\ref{1stAsyJ15}), and setting $d_{3}':=C\Gamma(-Y)[M^{Y}-(M^{*})^{Y}]$, lead to
\begin{align}
B_{1}(t)&=-\frac{1}{2}C\Gamma(-Y)\left[M^{Y}-\left(M^{*}\right)^{Y}\right]t^{\frac{1}{2}}+t^{1-\frac{Y}{2}}\left(t^{1-\frac{Y}{2}}d_{31}'+o(t^{1-\frac{Y}{2}})+\frac{C\sigma^{1-Y}}{2(Y-1)}\widetilde{\mathbb{E}}\left(\left|W_{1}^{*}\right|^{1-Y}\right)\right)+o(t^{\frac{1}{2}})\nonumber\\
\label{ExpB1} &=-\frac{1}{2}d_{3}'\,t^{\frac{1}{2}}+\frac{C\sigma^{1-Y}}{2(Y-1)}\widetilde{\mathbb{E}}\left(\left|W_{1}^{*}\right|^{1-Y}\right)t^{1-\frac{Y}{2}}+d_{31}'t^{2-Y}+o(t^{\frac{1}{2}})+o(t^{2-Y}),\quad t\rightarrow 0.
\end{align}

\noindent
\textbf{Step 2.} Next, we tackle $B_{2}(t)$ by decomposing it {as}
\begin{align}
B_{2}(t)&=\int_{0}^{\infty}\widetilde{\mathbb{E}}\!\left(\left(e^{-t^{\frac{1}{Y}}\!\widetilde{U}_{1}}\!-\!1\right)\!{\bf 1}_{\left\{Z_{1}\leq-t^{\frac{1}{2}-\frac{1}{Y}}\!w\right\}}\right)\frac{1\!-\!e^{-\sqrt{t}w}}{\sqrt{t}}\frac{e^{-\frac{w^{2}}{2\sigma^{2}}}}{\sqrt{2\pi\sigma^{2}}}\,dw+\int_{0}^{\infty}\widetilde{\mathbb{P}}\!\left(Z_{1}\!\leq\!-t^{\frac{1}{2}-\frac{1}{Y}}\!w\right)\frac{1\!-\!e^{-\sqrt{t}w}}{\sqrt{t}}\frac{e^{-\frac{w^{2}}{2\sigma^{2}}}}{\sqrt{2\pi\sigma^{2}}}\,dw\nonumber\\
\label{DfnC1C2} &=:B_{21}(t)+B_{22}(t).
\end{align}
We begin with proving that $B_{21}(t)=o(t^{1/2})$. To this end, set
\begin{align*}
B_{21}^{(1)}(t):=\int_{0}^{\infty}b_{21}^{(1)}(t;w)\frac{1-e^{-\sqrt{t}w}}{\sqrt{t}}\frac{e^{-\frac{w^{2}}{2\sigma^{2}}}}{\sqrt{2\pi\sigma^{2}}}\,dw,\quad
b_{21}^{(1)}(t;w):=\widetilde{\mathbb{E}}\left(\left(e^{-t^{\frac{1}{Y}}\widetilde{U}_{1}}-1\right){\bf 1}_{\{Z_{1}\leq -t^{\frac{1}{2}-\frac{1}{Y}}w,\,\widetilde{U}_{1}<0\}}\right).
\end{align*}
By (\ref{expUt}), for any $0<t<1$ and $w>0$,
\begin{align*}
0\leq t^{-\frac{1}{2}}b_{21}^{(1)}(t;w)&=t^{-\frac{1}{2}}\widetilde{\mathbb{E}}\left({\bf 1}_{\left\{Z_{1}\leq-t^{\frac{1}{2}-\frac{1}{Y}}w,\,\widetilde{U}_{1}<0\right\}}\int_{-\infty}^{0}{\bf 1}_{\{t^{\frac{1}{Y}}\widetilde{U}_{1}\leq u\leq 0\}}e^{-u}\,du\right)\\
&\leq t^{-\frac{1}{2}}\!\int_{-\infty}^{0}e^{-u}\widetilde{\mathbb{P}}\left(\widetilde{U}_{1}\leq t^{-\frac{1}{Y}}u\right)du\leq\widetilde{\mathbb{E}}\left(e^{-\widetilde{U}_{1}}\right)t^{-\frac{1}{2}}\!\int_{-\infty}^{0}e^{-u\left(1-t^{-\frac{1}{Y}}\right)}du=e^{\eta}\frac{t^{\frac{1}{Y}-\frac{1}{2}}}{1-t^{\frac{1}{Y}}}.
\end{align*}
Since $Y\in(1,2)$, dominated convergence implies that $B_{21}^{(1)}(t)=o(t^{1/2})$, as $t\rightarrow 0$. Next, consider
\begin{equation*}
B_{21}^{(2)}(t):=\int_{0}^{\infty}b_{21}^{(2)}(t;w)\frac{1-e^{-\sqrt{t}w}}{\sqrt{t}}\frac{e^{-\frac{w^{2}}{2\sigma^{2}}}}{\sqrt{2\pi\sigma^{2}}}\,dw,\quad
b_{21}^{(2)}(t;w):=\widetilde{\mathbb{E}}\left(\left(e^{-t^{\frac{1}{Y}}\widetilde{U}_{1}}-1\right){\bf 1}_{\{Z_{1}\leq -t^{\frac{1}{2}-\frac{1}{Y}}w,\,\widetilde{U}_{1}\geq 0\}}\right),
\end{equation*}
and further consider the decomposition
\begin{align*}
b_{21}^{(2)}(t;w)=\widetilde{\mathbb{E}}\left(\left(e^{-t^{\frac{1}{Y}}\widetilde{U}_{1}}-1+t^{\frac{1}{Y}}\widetilde{U}_{1}\right){\bf 1}_{\left\{Z_{1}\leq-t^{\frac{1}{2}-\frac{1}{Y}}w,\,\widetilde{U}_{1}\geq 0\right\}}\right)-t^{\frac{1}{Y}}\widetilde{\mathbb{E}}\left(\widetilde{U}_{1}{\bf 1}_{\left\{Z_{1}\leq-t^{\frac{1}{2}-\frac{1}{Y}}w,\,\widetilde{U}_{1}\geq 0\right\}}\right).
\end{align*}
Note that, as $t\rightarrow 0$,
\begin{align*}
0\leq t^{-\frac{1}{2}}\int_{0}^{\infty}t^{\frac{1}{Y}}\widetilde{\mathbb{E}}\left(\widetilde{U}_{1}{\bf 1}_{\left\{Z_{1}\leq-t^{\frac{1}{2}-\frac{1}{Y}}w,\,\widetilde{U}_{1}\geq 0\right\}}\right)\frac{1-e^{-\sqrt{t}w}}{\sqrt{t}}\frac{e^{-\frac{w^{2}}{2\sigma^{2}}}}{\sqrt{2\pi\sigma^{2}}}\,dw\leq t^{\frac{1}{Y}-\frac{1}{2}}\widetilde{\mathbb{E}}\left(|\widetilde{U}_{1}|\right)\int_{0}^{\infty}\frac{w\,e^{-\frac{w^{2}}{2\sigma^{2}}}}{\sqrt{2\pi\sigma^{2}}}\,dw\rightarrow 0.
\end{align*}
Moreover, by (\ref{UIn1Gen}) and the decomposition $\widetilde{U}_{t}=M^{*}\bar{U}_{t}^{(p)}-G^{*}\bar{U}_{t}^{(n)}$, for any $t>0$ and $w>0$,
\begin{align*}
0\leq\widetilde{\mathbb{E}}\left(\left(e^{-t^{\frac{1}{Y}}\widetilde{U}_{1}}-1+t^{\frac{1}{Y}}\widetilde{U}_{1}\right){\bf 1}_{\left\{Z_{1}\leq-t^{\frac{1}{2}-\frac{1}{Y}}w,\,\widetilde{U}_{1}\geq 0\right\}}\right)&=\int_{0}^{\infty}\left(1-e^{-u}\right)\widetilde{\mathbb{P}}\left(\widetilde{U}_{1}\geq t^{-\frac{1}{Y}}u,Z_{1}\leq-t^{\frac{1}{2}-\frac{1}{Y}}w\right)du\\
&\leq 2^{Y+1}K_{1}\left[\left(M^{*}\right)^{Y}+\left(G^{*}\right)^{Y}\right]t\int_{0}^{\infty}\left(1-e^{-u}\right)u^{-Y}du.
\end{align*}
Hence, by the dominated convergence theorem,
\begin{align*}
t^{-\frac{1}{2}}\int_{0}^{\infty}\widetilde{\mathbb{E}}\left(\left(e^{-t^{\frac{1}{Y}}\widetilde{U}_{1}}-1+t^{\frac{1}{Y}}\widetilde{U}_{1}\right){\bf 1}_{\left\{Z_{1}\leq-t^{\frac{1}{2}-\frac{1}{Y}}w,\,\widetilde{U}_{1}\geq 0\right\}}\right)\frac{1-e^{-\sqrt{t}w}}{\sqrt{t}}\frac{e^{-\frac{w^{2}}{2\sigma^{2}}}}{\sqrt{2\pi\sigma^{2}}}\,dw\rightarrow 0,\quad t\rightarrow 0.
\end{align*}
We then conclude that $B_{21}(t)=o(t^{1/2})$. To finish, we analyze $B_{22}(t)$ defined via (\ref{DfnC1C2}). To this end, let
\begin{align}
\widetilde{B}_{22}(t)&:=t^{\frac{Y}{2}-1}B_{22}(t)-\frac{C\sigma^{1-Y}}{2Y}\widetilde{\mathbb{E}}\left(\left|W_{1}^{*}\right|^{1-Y}\right)\nonumber\\
&\,={\Red t^{\frac{Y}{2}-1}\int_{0}^{\infty}\widetilde{\mathbb{E}}\left(\frac{1-e^{-\sqrt{t}\sigma W_{1}^{*}}-\sigma \sqrt{t}W_{1}^{*}}{\sqrt{t}}{\bf 1}_{\{0\leq\sigma W_{1}^{*}\leq t^{\frac{1}{Y}-\frac{1}{2}}z\}}\right)p_{Z}(1,z)\,dz}\nonumber\\
\label{DecomJ23} &\quad\,+{\Red t^{\frac{Y}{2}-1}\int_{0}^{\infty}\widetilde{\mathbb{E}}\left(\sigma W_{1}^{*}{\bf 1}_{\{0\leq\sigma W_{1}^{*}\leq t^{\frac{1}{Y}-\frac{1}{2}}z\}}\right)\left(p_{Z}(1,z)-Cz^{-Y-1}\right)dz}.
\end{align}
From the inequality $0\leq e^{-\sqrt{t}\sigma W_{1}^{*}}-1+\sigma\sqrt{t}W_{1}^{*}\leq\sigma^{2}t(W_{1}^{*})^{2}/2$, valid when $W_{1}^{*}\geq 0$, and the estimate (\ref{Asydenpz}), it is easy to see that the first term in (\ref{DecomJ23}) is of order {\Red $O(t^{1/2})$}. The second term in (\ref{DecomJ23}) is just ${\Red t^{Y/2-1}}R_{t^{(2-Y)/2Y}}^{(1)}$ and, thus, applying Lemma \ref{thm:ThirdOrderAsyCGMYBLemma}, we conclude that
\begin{align}\label{eq:AsyB22}
\lim_{t\rightarrow 0}t^{\frac{Y}{2}-1}\widetilde{B}_{22}(t)=-\frac{C^{2}(2Y-1)\cos^{2}\left(\frac{\pi Y}{2}\right)\Gamma^{2}(-Y)}{\sqrt{2\pi}\sigma^{2Y-1}}\widetilde{\mathbb{E}}\left(\left|W_{1}^{*}\right|^{2Y-2}\right){=:d_{32}'}.
\end{align}
Thus, by combining (\ref{DfnC1C2}) and (\ref{eq:AsyB22}),
\begin{align}\label{1stAsyJ23}		
B_{2}(t)=\frac{C\sigma^{1-Y}}{2Y}\widetilde{\mathbb{E}}\left(|W_{1}^{*}|^{1-Y}\right)t^{1-\frac{Y}{2}}+d_{32}'t^{2-Y}+o(t^{\frac{1}{2}})+o(t^{2-Y}),\quad t\rightarrow 0.
\end{align}
\textbf{Step 3.} {\Red Finally}, we study the behavior of $B_{3}(t)$ by further decomposing it {as
\begin{align*}
B_{3}(t)&=\int_{0}^{\infty}\widetilde{\mathbb{E}}\left(\left(e^{-t^{\frac{1}{Y}}\widetilde{U}_{1}}-1\right){\bf 1}_{\left\{Z_{1}\geq t^{\frac{1}{2}-\frac{1}{Y}}w\right\}}\right)\frac{1-e^{\sqrt{t}w}}{\sqrt{t}}\frac{e^{-\frac{w^{2}}{2\sigma^{2}}}}{\sqrt{2\pi\sigma^{2}}}\,dw+\int_{0}^{\infty}\widetilde{\mathbb{P}}\left(Z_{1}\geq t^{\frac{1}{2}-\frac{1}{Y}}w\right)\frac{1-e^{\sqrt{t}w}}{\sqrt{t}}\frac{e^{-\frac{w^{2}}{2\sigma^{2}}}}{\sqrt{2\pi\sigma^{2}}}\,dw\nonumber\\
&\quad+\int_{0}^{\infty}\widetilde{\mathbb{E}}\left({\bf 1}_{\left\{Z_{1}\geq t^{\frac{1}{2}-\frac{1}{Y}}w\right\}}\left(\frac{e^{-t^{\frac{1}{Y}}\widetilde{U}_{1}}-e^{-t^{\frac{1}{Y}}\left(Z_{1}+\widetilde{U}_{1}\right)}}{\sqrt{t}}-t^{\frac{1}{Y}-\frac{1}{2}}Z_{1}\right)\right)e^{\sqrt{t}w}\frac{e^{-\frac{w^{2}}{2\sigma^{2}}}}{\sqrt{2\pi\sigma^{2}}}\,dw\nonumber\\
&\quad+\int_{0}^{\infty}t^{\frac{1}{Y}-\frac{1}{2}}\widetilde{\mathbb{E}}\left(Z_{1}{\bf 1}_{\left\{Z_{1}\geq t^{\frac{1}{2}-\frac{1}{Y}}w\right\}}\right)e^{\sqrt{t}w}\frac{e^{-\frac{w^{2}}{2\sigma^{2}}}}{\sqrt{2\pi\sigma^{2}}}\,dw\nonumber\\
&=:B_{31}(t)+B_{32}(t)+B_{33}(t)+B_{34}(t).
\end{align*}
First, $B_{32}(t)$} is similar to $B_{22}(t)$ in (\ref{DfnC1C2}) and, thus, arguments similar to those leading to (\ref{eq:AsyB22}) imply that
\begin{align*}
B_{32}(t)=-\frac{C\sigma^{1-Y}}{2Y}\widetilde{\mathbb{E}}\left(|W_{1}^{*}|^{1-Y}\right)t^{1-\frac{Y}{2}}-d_{32}'t^{2-Y}+o(t^{2-Y}),\quad t\rightarrow 0.
\end{align*}
Similarly, $B_{34}(t)$ is similar to the term $B_{13}(t)$ introduced in (\ref{Eq:DfnC123}) and, thus, using arguments similar to those leading to (\ref{1stAsyJ15}), we have
\begin{align*}
B_{34}(t)=\frac{C\sigma^{1-Y}}{2(Y-1)}\widetilde{\mathbb{E}}\left(|W_{1}^{*}|^{1-Y}\right)t^{1-\frac{Y}{2}}+d_{31}'t^{2-Y}+o(t^{\frac{1}{2}})+o(t^{2-Y}),\quad t\rightarrow 0.
\end{align*}
Since $(-\bar{U}_{1}^{(n)},-\bar{U}_{1}^{(p)})\ed(\bar{U}_{1}^{(p)},\bar{U}_{1}^{(n)})$, $B_{31}(t)$ {\Red has a form similar to} $B_{21}(t)$ defined in (\ref{DfnC1C2}), but with the role of the parameters $M^{*}$ and $G^{*}$ reversed {\Red and $e^{-\sqrt{t}w}$ replaced by $e^{\sqrt{t}w}$}. Therefore, as for $B_{21}(t)$, we have that $B_{31}(t)=o(t^{1/2})$, as $t\rightarrow 0$. To finish, we further decompose $B_{33}(t)$ as follows:
\begin{align}\label{DecomJ33}
B_{33}(t)=t^{-\frac{1}{2}}\int_{0}^{\infty}{\left[\left(\int_{-\infty}^{0}+\int_{0}^{\infty}\right)\left(e^{-x}-1\right)P_{t}(w,x)\,dx\right]}e^{\sqrt{t}w}\frac{e^{-\frac{w^{2}}{2\sigma^{2}}}}{\sqrt{2\pi\sigma^{2}}}\,dw,
\end{align}
where
\begin{align}\label{Ptw}
P_{t}(w,x):=\widetilde{\mathbb{P}}\left(Z_{1}\geq t^{\frac{1}{2}-\frac{1}{Y}}w,\,\widetilde{U}_{1}\leq t^{-\frac{1}{Y}}x\leq Z_{1}+\widetilde{U}_{1}\right).
\end{align}
When $x<0$ and $w>0$, by (\ref{expUt}), $P_{t}(w,x)\leq e^{\eta}e^{t^{-1/Y}x}$ and, thus, for $0<t<1$ and some constant $K>0$,
\begin{align}\label{AsyJ331}
0&\leq\frac{1}{t}\int_{0}^{\infty}{\left(\int_{-\infty}^{0}\left(e^{-x}-1\right)P_{t}(w,x)\,dx\right)}e^{\sqrt{t}w}\frac{e^{-\frac{w^{2}}{2\sigma^{2}}}}{\sqrt{2\pi\sigma^{2}}}\,dw\leq Kt^{\frac{2}{Y}-1}e^{\eta}\int_{0}^{\infty}\,e^{w}\frac{e^{-\frac{w^{2}}{2\sigma^{2}}}}{\sqrt{2\pi\sigma^{2}}}\,dw\rightarrow 0,\quad\text{as }\,t\rightarrow 0.
\end{align}
For the integral $\int_{0}^{\infty}\int_{0}^{\infty}$ in (\ref{DecomJ33}), we show below (see Appendix \ref{proofB}) that
\begin{equation}\label{NdedLmt2}
\lim_{t\rightarrow 0}\frac{1}{t}P_{t}(w,x)=\frac{C}{Y}\left[M^{Y}-\left(M^{*}\right)^{Y}\right]x^{-Y}.
\end{equation}
Moreover, by arguments similar to those leading to (\ref{UIn1Gen}), $t^{-1}\widetilde{\mathbb{P}}({\Red t^{-1/Y}}x\leq Z_{1}+\widetilde{U}_{1})\leq{\lambda}x^{-Y}$, for any $x>0$ and some constant {$\lambda>0$, and thus}, by the dominated convergence theorem,
\begin{align*}
\lim_{t\rightarrow 0}t^{-1}\int_{0}^{\infty}{\left(\int_{0}^{\infty}\left(e^{-x}-1\right)P_{t}(w,x)dx\right)}e^{\sqrt{t}w}\frac{e^{-\frac{w^{2}}{2\sigma^{2}}}}{\sqrt{2\pi\sigma^{2}}}\,dw=-\frac{C\Gamma(-Y)}{2}\left[M^{Y}-\left(M^{*}\right)^{Y}\right]=:d_{3}'.
\end{align*}
The above limit, together with (\ref{AsyJ331}), implies that {\Red $B_{33}(t)=t^{1/2}d_{3}'+o(t^{1/2})$} and, thus,
\begin{align}\label{eq:AsyB3}
B_{3}(t)&={\Red d'_{3}\,t^{\frac{1}{2}}}+\frac{C\sigma^{1-Y}}{2Y(Y-1)}\widetilde{\mathbb{E}}\left(|W_{1}^{*}|^{1-Y}\right)t^{1-\frac{Y}{2}}+{\left(d_{31}'-d_{32}'\right)t^{2-Y}}+o(t^{\frac{1}{2}})+{o(t^{2-Y}),\quad t\rightarrow 0.}
\end{align}
Finally, combining (\ref{DecomBt}), (\ref{DecomB1}), (\ref{ExpB1}), (\ref{1stAsyJ23}), and (\ref{eq:AsyB3}), we establish that
\begin{align*}
{\Delta_{0}(t)=\left(\frac{\tilde\gamma}{2}-\frac{\sigma^{2}}{4}+2d'_{3}\right)t^{\frac{1}{2}}+\frac{C\sigma^{1-Y}}{Y(Y-1)}\widetilde{\mathbb{E}}\left(\left|W^{*}_{1}\right|^{1-Y}\right)t^{1-\frac{Y}{2}}+{2\left(d_{31}'-d_{32}'\right)t^{2-Y}}+o(t^{\frac{1}{2}})+{o(t^{2-Y}),}}
\end{align*}
which yields (\ref{3rdAsyCGMYB1}), by noting that the coefficient of the first term above reduces to the expression $d_{31}$ in (\ref{eq:3rdCoefGenCGMY1}) and that $d_{32}=2(d_{31}'-d_{32}')$. \hfill $\Box$

\medskip
\noindent
\textbf{Proof of Theorem \ref{thm:ThirdOrderAsyPureCGMY}.}
{\Red In the pure-jump case, the decomposition \eqref{eq:DecomOPT} still holds. For $G_{1}(t)$ given by \eqref{eq:DefG1G2}, when $\kappa_{t}\neq 0$, changing variables to $w=v/\kappa_{t}$ leads to
\begin{align*}
G_{1}(t)=\kappa_{t}\int_{0}^{1}e^{-\kappa_{t}w}\mathbb{P}^{*}\left(t^{-1/Y}X_{t}\geq t^{-1/Y}\kappa_{t}w\right)dw=\kappa_{t}\int_{0}^{1}\mathbb{P}^{*}\left(Z\geq t^{-1/Y}\kappa_{t}w\right)dw+o(\kappa_{t}),\quad t\rightarrow 0,
\end{align*}
where $Z$ is a symmetric strictly $Y$-stable random variable under $\mathbb{P}^{*}$. In the last equality above, we have used the fact (cf.~\cite[Theorem 3.1]{Rosinski:2007}) that $t^{-1/Y}X_{t}$ converges in distribution to a symmetric strictly $Y$-stable random variable under $\mathbb{P}^{*}$, and that pointwise convergence of a sequence of distribution functions to a continuous distribution functions implies uniform convergence.

Now, we study the asymptotic behavior of $G_{2}(t)$ given by \eqref{eq:DefG1G2}.} Set {\Red $\tilde{\gamma}_{t}:=t^{1-1/Y}\tilde{\gamma}$} and $\vartheta:=-C\Gamma(-Y)[M^{Y}+(G^{*})^{Y}]$ and note that, in view of (\ref{Cent}) and (\ref{eta}), {\Red $d_{2}=\vartheta+\eta+\tilde{\gamma}/2$}. For future reference, it is also convenient to write $\vartheta$ as
\begin{equation}\label{UsfDcmp}
\vartheta=\frac{C}{Y}\left[M^{Y}+\left(G^{*}\right)^{Y}\right]\int_{0}^{\infty}\frac{e^{-t^{\frac{1}{Y}}v}-1}{t^{1-\frac{1}{Y}}}v^{-Y}dv,
\end{equation}
which follows from the well-known identity $\Gamma(1-Y)=\int_{0}^{\infty}\left(e^{-y}-1\right)y^{-Y}dy$ (see (14.18) in~\cite{Sato:1999}). Also, note that
\begin{align}\label{expUt} \widetilde{\mathbb{E}}\left(e^{-\widetilde{U}_{t}}\right)=\widetilde{\mathbb{E}}\left(e^{-t^{1/Y}\widetilde{U}_{1}}\right)=\widetilde{\mathbb{E}}\left(e^{-t^{1/Y}M^{*}\bar{U}^{(p)}_{1}}\right)\widetilde{\mathbb{E}}\left(e^{t^{1/Y}G^{*}\bar{U}^{(n)}_{1}}\right)=e^{\eta t},\quad{t\geq 0}.
\end{align}
From (\ref{DSMBoth}), (\ref{DcmLL}), (\ref{RX}), {\Red (\ref{eq:DefG1G2}),} and (\ref{expUt}), we have
\begin{align*}
{\Red G_{2}(t)=\mathbb{E}^{*}\left(1-e^{X_{t}^{+}}\right)=e^{-\eta t}\,\widetilde{\mathbb{E}}\left(e^{-\widetilde{U}_{t}}\left(1-e^{-X_{t}^{+}}\right)\right)=1-e^{-\eta t}\,\widetilde{\mathbb{E}}\left(e^{-\widetilde{U}_{t}-X_{t}^{+}}\right).}
\end{align*}
Set
\begin{align*}
\Delta_{1}(t):=t^{-\frac{1}{Y}}\,\widetilde{\mathbb{E}}\left(1-e^{-\left(\widetilde{U}_{t}+X_{t}^{+}\right)}-\left(\widetilde{U}_{t}+X_{t}^{+}\right)\right),\quad\Delta_{2}(t):=t^{-\frac{1}{Y}}\left(\widetilde{\mathbb{E}}\left(X_{t}^{+}\right)-\widetilde{\mathbb{E}}\left(Z_{t}^{+}\right)\right).
\end{align*}
Then, recalling that $\widetilde{\mathbb{E}}\,(\widetilde{U}_{t})=0$ and $\widetilde{\mathbb{E}}\left(Z_{t}^{+}\right)=t^{1/Y}\widetilde{\mathbb{E}}{\left(Z_{1}^{+}\right)}$, we have the decomposition
\begin{align}
A(t)&:=t^{\frac{1}{Y}-1}\left(t^{-\frac{1}{Y}}{\Red G_{2}(t)}-\widetilde{\mathbb{E}}(Z_{1}^{+})\right)-d_{2}\nonumber\\
&\,=\left(t^{\frac{1}{Y}-1}\Delta_{1}(t)-\vartheta\right)+\left(t^{\frac{1}{Y}-1}\Delta_{2}(t)-{\Red \frac{\tilde{\gamma}}{2}}\right)-\frac{e^{-\eta t}-1+\eta t}{t}\,\widetilde{\mathbb{E}}\left(e^{-\widetilde{U}_{t}-X_{t}^{+}}\right)-\eta t^{\frac{1}{Y}}\Delta_{1}(t)-\eta\,\widetilde{\mathbb{E}}\left(X_{t}^{+}\right)\nonumber\\
\label{eq:DecomA}
&\,=:A_{1}(t)+A_{2}(t)+A_{3}(t)-A_{4}(t)-A_{5}(t).
\end{align}
We will prove that $A_{1}(t)={\Red O(t^{2/Y-1})}$ (and so ${\Red t^{1/Y}}\Delta_{1}(t)=O(t)$), and that $A_{2}={\Red O(t^{1-1/Y})}$. These results, in turn, imply that $A_{i}(t)=O(t)={\Red o(t^{2/Y-1})}={\Red o(t^{1-1/Y})}$, $i=3,4$, and that $A_{5}(t)={\Red O(t^{1/Y})}={\Red o(t^{2/Y-1})}={\Red o(t^{1-1/Y})}$, since $\kappa_{t}t^{-1/Y}\rightarrow 0$. So, it remains to analyze the asymptotic behavior of $A_{1}(t)$ and $A_{2}(t)$. This is done in two steps.

\smallskip
\noindent
\textbf{Step 1.} By Fubini's theorem and a change of variables,
\begin{align}	
A_{1}(t)&=\left(\int_{0}^{\infty}\frac{e^{-t^{\frac{1}{Y}}v}-1}{t^{1-\frac{1}{Y}}}\widetilde{\mathbb{P}}\left({t^{-\frac{1}{Y}}\left(X_{t}^{+}+\widetilde{U}_{t}\right)}\geq v\right)dv-\vartheta\right)-\int_{0}^{\infty}\frac{e^{t^{\frac{1}{Y}}v}-1}{t^{1-\frac{1}{Y}}}\widetilde{\mathbb{P}}\left({t^{-\frac{1}{Y}}\left(X_{t}^{+}+\widetilde{U}_{t}\right)}\leq-v\right)dv\nonumber\\
\label{eq:AsyA145} &{=:B_{1}(t)-B_{2}(t)}.
\end{align}
For $B_{2}(t)$, using the decompositions (\ref{DcmLL})-(\ref{RX}) as well as the self-similarity of $((Z_{t},\widetilde{U}_{t}))_{t\geq 0}$,
\begin{align}\label{eq:AsyA3}
\lim_{t\rightarrow 0}t^{1-\frac{2}{Y}}B_{2}(t)=\lim_{t\rightarrow 0}\int_{0}^{\infty}\frac{e^{t^{\frac{1}{Y}}v}-1}{t^{\frac{1}{Y}}}\,\widetilde{\mathbb{P}}\left(\left(Z_{1}+\tilde{\gamma}_{t}\right)^{+}+\widetilde{U}_{1}\leq-v\right)dv=\int_{0}^{\infty}v\,\widetilde{\mathbb{P}}\left(Z_{1}^{+}+\widetilde{U}_{1}\leq -v\right)dv,
\end{align}
where the second equality follows from the dominated convergence theorem, which applies in view of the following direct consequences of (\ref{expUt}):
\begin{align*}
\frac{e^{t^{\frac{1}{Y}}\!v}-1}{t^{\frac{1}{Y}}}\,\widetilde{\mathbb{P}}\left(\left(Z_{1}+\tilde\gamma_{t}\right)^{+}+\widetilde{U}_{1}\leq -v\right)\leq ve^{t^{\frac{1}{Y}}v}e^{-v}\,\widetilde{\mathbb{E}}\left(e^{-\widetilde{U}_{1}}\right)= e^{\eta}v\,e^{\left(t^{\frac{1}{Y}}-1\right)v}\leq e^{\eta}v\,e^{-v/2}.
\end{align*}
To analyze $B_{1}(t)$,  we again use the decompositions (\ref{DcmLL})-(\ref{RX}) as well as the self-similarity of $((Z_{t},\widetilde{U}_{t}))_{t\geq 0}$ to get
\begin{align}
t^{1-\frac{2}{Y}}B_{1}(t)&\,=\int_{0}^{\infty}\frac{e^{-t^{\frac{1}{Y}}v}-1}{t^{\frac{1}{Y}}}\left(\widetilde{\mathbb{P}}\left(Z_{1}+\tilde{\gamma}_{t}>0,\,Z_{1}+\tilde{\gamma}_{t}+\widetilde{U}_{1}\geq v\right)-\frac{CM^{Y}}{Yv^{Y}}\right)dv\nonumber\\	
\label{eq:DecomA2}
&\quad+\int_{0}^{\infty}\frac{e^{-t^{\frac{1}{Y}}v}-1}{t^{\frac{1}{Y}}}\left(\widetilde{\mathbb{P}}\left(Z_{1}+\tilde{\gamma}_{t}\leq{}0,\,{\widetilde{U}_{1}}\geq v\right)-\frac{C(G^{*})^{Y}}{Yv^{Y}}\right)dv,
\end{align}
where we have used (\ref{UsfDcmp}). As suggested from the previous decomposition, the limit of each of the terms therein can be obtained by passing $\lim_{t\rightarrow 0}$ into the various integrals to get
\begin{align}\label{NdTp1}
\lim_{t\rightarrow 0}t^{1-\frac{2}{Y}}B_{1}(t)=-\int_{0}^{\infty}v\left(\widetilde{\mathbb{P}}\left(Z_{1}^{+}+\widetilde{U}_{1}\geq v\right)-\frac{CM^{Y}}{Yv^{Y}}-\frac{C(G^{*})^{Y}}{Yv^{Y}}\right)dv.
\end{align}
For the sake of a more streamlined proof, we defer the justification of the latter operation to Appendix \ref{proofB}. Combining (\ref{eq:AsyA145}), (\ref{eq:AsyA3}), and (\ref{NdTp1}), we obtain that
\begin{align}\label{ConclStep1}
\lim_{t\rightarrow 0}t^{1-\frac{2}{Y}}A_{1}(t)=-\int_{0}^{\infty}v\left(\widetilde{\mathbb{P}}\left(Z_{1}^{+}+\widetilde{U}_{1}\geq v\right)-\frac{CM^{Y}}{Yv^{Y}}-\frac{C(G^{*})^{Y}}{Yv^{Y}}\right)dv-\frac{1}{2}\widetilde{\mathbb{E}}\left(\left(\left(Z_{1}^{+}+\widetilde{U}_{1}\right)^{-}\right)^{2}\right)=:d_{32}.
\end{align}

\noindent
\textbf{Step 2.} Now, we analyze the behavior of $A_{2}={\Red t^{1/Y-1}}\Delta_{2}(t)-\tilde{\gamma}/2$. By the self-similarity of $(Z_{t})_{t\geq 0}$,
\begin{align*}
\Delta_{2}(t)=\widetilde{\mathbb{E}}\left(\left(Z_{1}+\tilde\gamma_{t}\right)^{+}-Z_{1}^{+}\right)=\int_{0}^{\infty}\left(\widetilde{\mathbb{P}}\left(Z_{1}\geq u-\tilde{\gamma}_{t}\right)-\widetilde{\mathbb{P}}\left(Z_{1}\geq u\right)\right)du=\int_{0}^{\infty}\int_{u-\tilde{\gamma}_{t}}^{u}p_{Z}(w)\,dw\,du,
\end{align*}
where for simplicity we wrote $p_{Z}(u)$ for the density $p_{Z}(1,u)$ of $Z_{1}$. From the symmetry of $Z_{1}$, {\Red $\tilde{\gamma}/2=\tilde{\gamma}\int_{0}^{\infty}p_{Z}(u)du$} and, thus, recalling that {\Red $\tilde{\gamma}_{t}:=t^{1-1/Y}\tilde\gamma$,
\begin{align}
\lim_{t\rightarrow 0}t^{\frac{1}{Y}-1}A_{2}(t)&=\lim_{t\rightarrow 0}t^{\frac{1}{Y}-1}\left(t^{\frac{1}{Y}-1}\tilde{\gamma_{t}}\int_{0}^{\infty}\left(\frac{1}{\tilde{\gamma}_{t}}\int_{u-\tilde{\gamma}_{t}}^{u}p_{Z}(w)\,dw-p_{Z}(u)\right)du\right)\nonumber\\
&=\lim_{t\rightarrow 0}t^{\frac{2}{Y}-2}\,\tilde{\gamma}_{t}^{2}\int_{-1}^{0}v\left(\int_{0}^{1}\int_{0}^{\infty}p'_{Z}(u+\beta\tilde{\gamma}_{t}v)\,du\,d\beta\right)dv\nonumber\\
\label{eq:AsytildeA2} &=-\lim_{t\rightarrow 0}t^{\frac{2}{Y}-2}\tilde{\gamma}^{2}_{t}\int_{-1}^{0}v\left(\int_{0}^{1}p_{Z}\left(\beta\tilde{\gamma}vt^{1-\frac{1}{Y}}\right)d\beta\right)dv=\frac{\tilde{\gamma}^{2}p_{Z}(0)}{2}=:d_{31}.
\end{align}
The} expression for $d_{31}$ as given in (\ref{eq:3rdCoefPureCGMY1}) is obtained from the power series representation of $p_{Z}$ around $z=0$ shown, for example, in (14.30) of~\cite{Sato:1999}. Finally, combining (\ref{ConclStep1}) and (\ref{eq:AsytildeA2}) with (\ref{eq:DecomA}), we obtain (\ref{eq:3rdExpPureCGMY1}).\hfill $\Box$

\section{Further Proofs}\label{proofB}

\medskip
\noindent
\textbf{Proof of Lemma~\ref{Bnd1TailSt}.}
From the leading term in the expansion (\ref{eq:2ndAsyStableTailDist}), there exists $N>0$ such that, for any $x>0$,
\begin{align*}
\widetilde{\mathbb{P}}\left(\bar{U}_{1}^{(p)}\geq x\right)&=\widetilde{\mathbb{P}}\left(\bar{U}_{1}^{(p)}\geq x\right)\left({\bf 1}_{\{x\geq N\}}+{\bf 1}_{\{x<N\}}\right)\leq \frac{2C}{Y}x^{-Y}{\bf 1}_{\{x\geq N\}}+\frac{N^{Y}}{x^{Y}}{\bf 1}_{\{x<N\}}\leq\left(2CY^{-1}+N^{Y}\right)x^{-Y},
\end{align*}
and the first relationship in (\ref{UIn1Gen}) follows by setting $K_{1}=2CY^{-1}+N^{Y}$. Similarly, from (\ref{eq:2ndAsyStableTailDist}), there exists $N>0$ such that, for any $x>0$,
\begin{align*}
\left|\widetilde{\mathbb{P}}\left(\bar{U}_{1}^{(p)}\geq x\right)-\frac{C}{Y}x^{-Y}\right|&=\left|\widetilde{\mathbb{P}}\left(\bar{U}_{1}^{(p)}\geq x\right)-\frac{C}{Y}x^{-Y}\right|\left({\bf 1}_{\{x\geq N\}}+{\bf 1}_{\{x<N\}}\right)\\
&\leq\frac{C^{2}}{\pi}\left|\sin(2\pi Y)\right|\Gamma(2Y)\Gamma^{2}(-Y)x^{-2Y}{\bf 1}_{\{x\geq N\}}+\left(\widetilde{\mathbb{P}}\left(\bar{U}_{1}^{(p)}\geq x\right)+\frac{C}{Y}x^{-Y}\right){\bf 1}_{\{x<N\}}\\
&\leq\left(\frac{C^{2}}{\pi}\left|\sin(2\pi Y)\right|\Gamma(2Y)\Gamma^{2}(-Y)+N^{2Y}+CN^{Y}Y^{-1}\right)x^{-2Y}.
\end{align*}
The second identity in (\ref{UIn1Gen}) follows by setting $K_{2}=C^{2}\left|\sin(2\pi Y)\right|\Gamma(2Y)\Gamma^{2}(-Y)/\pi+N^{2Y}+CN^{Y}Y^{-1}$.\hfill $\Box$

\medskip
\noindent
\textbf{Proof of (\ref{NdTp1}).} We begin with $B_{11}(t)$. Using (\ref{DcmLL}) and (\ref{RX}), leads to the decomposition
\begin{align*}
&\widetilde{\mathbb{P}}\left(Z_{1}+\tilde{\gamma}_{t}>0,\,Z_{1}+\tilde{\gamma}_{t}+\widetilde{U}_{1}\geq v\right)-\frac{CM^{Y}}{Yv^{Y}}\\
&\quad=\widetilde{\mathbb{P}}\left(\bar{U}_{1}^{(p)}+\tilde{\gamma}_{t}\geq -\bar{U}_{1}^{(n)}\geq\frac{v+M^{*}\tilde{\gamma}_{t}}{M+G}\right)+\widetilde{\mathbb{P}}\left(\bar{U}_{1}^{(p)}\geq\frac{v+G\bar{U}_{1}^{(n)}-\tilde{\gamma}_{t}}{M},\,-\bar{U}_{1}^{(n)}<\frac{v+M^{*}\tilde{\gamma}_{t}}{M+G}\right)-\frac{CM^{Y}}{Yv^{Y}}\\
&\quad=:b_{11}^{(1)}(t;v)+b_{11}^{(2)}(t;v).
\end{align*}
For any $v>0$ and $t$ small enough (so that $G^{*}|\tilde\gamma_{t}|<1$ and $M^{*}|\tilde\gamma_{t}|<1$),
\begin{align*}
b_{11}^{(1)}(t;v)\leq v\widetilde{\mathbb{P}}\left(\bar{U}_{1}^{(p)}\geq\frac{v\!+\!M^{*}\tilde{\gamma}_{t}}{M\!+\!G}-\tilde{\gamma}_{t}\right)\widetilde{\mathbb{P}}\left(-\bar{U}_{1}^{(n)}\geq\frac{v\!+\!M^{*}\tilde{\gamma}_{t}}{M\!+\!G}\right)\leq v{\bf 1}_{\{v\leq 1\}}+v{\bf 1}_{\{v>1\}}\min\left(1,K_{1}^{2}(M\!+\!G)^{2Y}v^{-2Y}\right),
\end{align*}
where $K_{1}\in(0,\infty)$ is given as in (\ref{UIn1Gen}). We now consider {$b_{11}^{(2)}(t;v)$}. It suffices to consider $v>1$, since $|b_{11}^{(2)}(t;v)|\leq v(1+CY^{-1}M^{Y}v^{-Y})$, which is integrable on $\{v\leq 1\}$. We also let $t$ be small enough, so that $|\tilde{\gamma}_{t}|<1$, $G^{*}|\tilde{\gamma}_{t}|<1$, and $M^{*}|\tilde{\gamma}_{t}|<1$. Then, for any $v>1$,
\begin{align*}
\left|b_{11}^{(2)}(t;v)\right|&\leq v\int_{-\infty}^{\frac{v+M^{*}\tilde{\gamma}_{t}}{M+G}}p_{U}(1,y)\left|\widetilde{\mathbb{P}}\left(\bar{U}_{1}^{(p)}\geq\frac{v-Gy-\tilde{\gamma}_{t}}{M}\right)-\frac{CM^{Y}}{Y(v-Gy-\tilde{\gamma}_{t})^{Y}}\right|dy\nonumber\\
&\quad+v\int_{-\infty}^{\frac{v+M^{*}\tilde{\gamma}_{t}}{M+G}}p_{U}(1,y)\frac{CM^{Y}}{Y}\left|(v-Gy-\tilde{\gamma}_{t})^{-Y}-v^{-Y}\right|dy +\frac{CM^{Y}}{Yv^{Y-1}}\widetilde{\mathbb{P}}\left(\bar{U}_{1}^{(p)}\geq\frac{v+M^{*}\tilde{\gamma}_{t}}{M+G}\right)\nonumber\\
&{=:D_{t}^{(1)}(v)+D_{t}^{(2)}(v)+D_{t}^{(3)}(v)}.
\end{align*}
Next, by (\ref{UIn1Gen}), we have $D_{t}^{(1)}(v)\leq K_{2}(M+G)^{2Y}v^{1-2Y}$, for any $v>1$. Using the convexity and monotonicity of the function $f(x)=x^{-Y}$ on $(0,\infty)$, $D_{t}^{(2)}(v)\leq CM^{Y}v^{-Y}(G\widetilde{\mathbb{E}}|\bar{U}_{1}^{(p)}|+1)$. Finally, again by (\ref{UIn1Gen}), we have $D_{t}^{(3)}(v)\leq K_{1}CM^{Y}Y^{-1}v^{1-2Y}$, for any $v>1$. Combining the previous estimates, it is now clear that we can apply the dominated convergence theorem {\Red to the first integral in (\ref{eq:DecomA2})} to obtain its limit as $t\rightarrow 0$. One can apply similar arguments to justify passing the limit {\Red in} the {\Red second} integral in (\ref{eq:DecomA2}).\hfill $\Box$

\medskip
\noindent
\textbf{Proof of (\ref{AsympDescC2}).} First, change variables, $x={\Red t^{1/Y}}u$, in the integral of the term $B_{12}(t)$ defined in (\ref{Eq:DfnC123}), so that
\begin{align*}	
B_{12}(t)=t^{-\frac{1}{2}}e^{-(\eta t+\sqrt{t}\tilde{\gamma}_{t})}\int_{0}^{\infty}\left(\int_{\mathbb{R}}\left(e^{-x}-1\right)\widetilde{\mathbb{P}}\left(-t^{\frac{1}{2}-\frac{1}{Y}}w\leq Z_{1}\leq 0,\,\widetilde{U}_{1}+Z_{1}\leq t^{-\frac{1}{Y}}x\leq\widetilde{U}_{1}\right)dx\right)e^{-\sqrt{t}w}\frac{e^{-\frac{w^{2}}{2\sigma^{2}}}}{\sqrt{2\pi\sigma^{2}}}\,dw.
\end{align*}
We next prove that $B_{12}(t)=o(t^{1/2})$ as $t\rightarrow 0$. To this end, let
\begin{align*}
B_{12}^{(1)}(t)&=\int_{0}^{\infty}\left[\int_{0}^{\infty}\left(1-e^{-x}\right)\widetilde{\mathbb{P}}\left(-t^{\frac{1}{2}-\frac{1}{Y}}w\leq Z_{1}\leq 0,\,\widetilde{U}_{1}+Z_{1}\leq t^{-\frac{1}{Y}}x\leq\widetilde{U}_{1}\right)dx\right]e^{-\sqrt{t}w}\frac{e^{-\frac{w^{2}}{2\sigma^{2}}}}{\sqrt{2\pi\sigma^{2}}}\,dw,\\
B_{12}^{(2)}(t)&=\int_{0}^{\infty}\left[\int_{-\infty}^{0}\left(e^{-x}-1\right)\widetilde{\mathbb{P}}\left(-t^{\frac{1}{2}-\frac{1}{Y}}w\leq Z_{1}\leq 0,\,\widetilde{U}_{1}+Z_{1}\leq t^{-\frac{1}{Y}}x\leq\widetilde{U}_{1}\right)dx\right]e^{-\sqrt{t}w}\frac{e^{-\frac{w^{2}}{2\sigma^{2}}}}{\sqrt{2\pi\sigma^{2}}}\,dw.
\end{align*}
For any $t>0$, $w>0$ and $x>0$, by (\ref{UIn1Gen}) and (\ref{Ptw}),
\begin{align*}
\frac{1}{t}P_{t}(x,w)&=\widetilde{\mathbb{P}}\left(-t^{\frac{1}{2}-\frac{1}{Y}}w\leq\bar{U}_{1}^{(p)}+\bar{U}_{1}^{(n)}\leq 0,\,\left(M^{*}+1\right)\bar{U}_{1}^{(p)}-\left(G^{*}-1\right)\bar{U}_{1}^{(n)}\leq t^{-\frac{1}{Y}}x\leq M^{*}\bar{U}_{1}^{(p)}-G^{*}\bar{U}_{1}^{(n)}\right)\\
&\leq\frac{1}{t}\,\widetilde{\mathbb{P}}\left(\frac{t^{-\frac{1}{Y}}{x}}{M^{*}+G^{*}}\leq -\bar{U}_{1}^{(n)}\leq\frac{t^{-\frac{1}{Y}}x+\left(M^{*}+1\right)t^{\frac{1}{2}-\frac{1}{Y}}w}{M^{*}+G^{*}},\,\frac{t^{-\frac{1}{Y}}x+G^{*}\bar{U}^{(n)}_{1}}{M^{*}}\leq\bar{U}_{1}^{(p)}\right)\\
&\leq K_{1}\left(M^{*}+G^{*}\right)^{Y}x^{-Y}\,\widetilde{\mathbb{P}}\left(\frac{t^{-\frac{1}{Y}}M^{*}x-G^{*}\left(M^{*}+1\right)t^{\frac{1}{2}-\frac{1}{Y}}w}{M^{*}\left(M^{*}+G^{*}\right)}\leq\bar{U}_{1}^{(p)}\right)\rightarrow 0,\quad t\rightarrow 0,
\end{align*}
while for $t>0$, $w>0$ and $x<0$,
\begin{align*} \frac{1}{t}P_{t}(x,w)\leq\frac{2}{t}\widetilde{\mathbb{P}}\left(\bar{U}_{1}^{(p)}\leq\frac{t^{-\frac{1}{Y}}x}{2\left(M^{*}+G^{*}\right)}\right)\leq\frac{2}{t}\widetilde{\mathbb{E}}\left(e^{-\bar{U}_{1}^{(p)}}\right)\exp\left(\frac{t^{-\frac{1}{Y}}x}{2\left(M^{*}+G^{*}\right)}\right)\rightarrow 0,\quad t\rightarrow 0.
\end{align*}
It follows from dominated convergence that $B_{12}^{(1)}(t)=o(t)$ and $B_{12}^{(2)}(t)=o(t)$, which in turn implies (\ref{AsympDescC2}).\hfill $\Box$

\medskip
\noindent
\textbf{Proof of (\ref{NdedLmt2}).} First, for any $t>0$, $x>0$ and $w>0$, by (\ref{Ptw}),
\begin{align*}
\frac{1}{t}P_{t}(w,x) &=\frac{1}{t}\widetilde{\mathbb{P}}\left(\bar{U}_{1}^{(p)}+\bar{U}_{1}^{(n)}\geq t^{\frac{1}{2}-\frac{1}{Y}}w,\,M^{*}\bar{U}_{1}^{(p)}-G^{*}\bar{U}_{1}^{(n)}\leq t^{-\frac{1}{Y}}x\leq M\bar{U}_{1}^{(p)}-G\bar{U}_{1}^{(n)}\right)\\
&=\frac{1}{t}\int_{\mathbb{R}}\widetilde{\mathbb{P}}\left(\bar{U}_{1}^{(p)}\geq t^{\frac{1}{2}-\frac{1}{Y}}w+u,\,\frac{t^{-\frac{1}{Y}}x-Gu}{M}\leq\bar{U}_{1}^{(p)}\leq\frac{t^{-\frac{1}{Y}}x-G^{*}u}{M^{*}}\right)p_{U}(1,u)\,du.
\end{align*}
Note that
\begin{align*}
&\frac{t^{-\frac{1}{Y}}x-Gu}{M}\leq\frac{t^{-\frac{1}{Y}}x-G^{*}u}{M^{*}}\,\,\Leftrightarrow\,\,u\leq\frac{t^{-\frac{1}{Y}}x}{M+G},\quad t^{\frac{1}{2}-\frac{1}{Y}}w+u\leq\frac{t^{-\frac{1}{Y}}x-G^{*}u}{M^{*}}\,\,\Leftrightarrow\,\,u\leq\frac{t^{-\frac{1}{Y}}x-M^{*}t^{\frac{1}{2}-\frac{1}{Y}}w}{M+G},\\
&t^{\frac{1}{2}-\frac{1}{Y}}w+u\leq\frac{t^{-\frac{1}{Y}}x-Gu}{M}\,\,\Leftrightarrow\,\,u\leq\frac{t^{-\frac{1}{Y}}x-Mt^{\frac{1}{2}-\frac{1}{Y}}w}{M+G}.
\end{align*}
Hence,
\begin{align*}
\frac{1}{t}P_{t}(w,x)&=\frac{1}{t}\int_{-\infty}^{\frac{t^{-\frac{1}{Y}}x-Mt^{\frac{1}{2}-\frac{1}{Y}}w}{M+G}}\widetilde{\mathbb{P}}\left(\frac{t^{-\frac{1}{Y}}x-Gu}{M}\leq\bar{U}_{1}^{(p)}\leq\frac{t^{-\frac{1}{Y}}x-G^{*}u}{M^{*}}\right)p_{U}(1,u)\,du\nonumber\\
&\quad+\frac{1}{t}\int_{\frac{t^{-\frac{1}{Y}}x-Mt^{\frac{1}{2}-\frac{1}{Y}}w}{M+G}}^{\frac{t^{-\frac{1}{Y}}x-M^{*}t^{\frac{1}{2}-\frac{1}{Y}}w}{M+G}}\widetilde{\mathbb{P}}\left(t^{\frac{1}{2}-\frac{1}{Y}}w+u\leq\bar{U}_{1}^{(p)}\leq\frac{t^{-\frac{1}{Y}}x-G^{*}u}{M^{*}}\right)p_{U}(1,u)\,du\nonumber\\
&=:I_{1}(t;w,x)+I_{2}(t;w,x).
\end{align*}
For $I_{1}(t;w,x)$, note that for any $t>0$, $x>0$ and $w>0$,
\begin{align*}
u\leq\frac{t^{-\frac{1}{Y}}x-Mt^{-\frac{1}{Y}}w}{M+G}<\frac{t^{-\frac{1}{Y}}x}{M+G}<\frac{t^{-\frac{1}{Y}}x}{G}\,\,\Rightarrow\,\,t^{-\frac{1}{Y}}x-Gu>0,\quad\frac{x-Mw\sqrt{t}}{M+G}>0\,\,\Leftrightarrow\,\,t<\frac{x^{2}}{M^{2}w^{2}}.
\end{align*}
Hence, by (\ref{eq:2ndAsyStableTailDist}) and the dominated convergence theorem, for any $x>0$, $w>0$ and $u\leq {\Red t^{-1/Y}}(x-Mw)/(M+G)$,
\begin{align*}
\lim_{t\rightarrow 0}I_{1}(t;w,x)&=\int_{\mathbb{R}}p_{U}(1,u)\left(\lim_{t\rightarrow 0}\frac{1}{t}\widetilde{\mathbb{P}}\!\left(\frac{t^{-\frac{1}{Y}}x-Gu}{M}\leq\bar{U}_{1}^{+}\leq\frac{t^{-\frac{1}{Y}}x-G^{*}u}{M^{*}}\right){\bf 1}_{\left\{u\leq\frac{t^{-1/Y}x-Mt^{{1/2-1/Y}}w}{M+G}\right\}}\right)du\nonumber\\
&=\int_{\mathbb{R}}p_{U}(1,u)\!\left(\lim_{t\rightarrow 0}\frac{1}{t}\widetilde{\mathbb{P}}\left(\bar{U}_{1}^{+}\geq\frac{t^{-\frac{1}{Y}}x-Gu}{M}\right)\!\right)du-\int_{\mathbb{R}}p_{U}(1,u)\!\left(\lim_{t\rightarrow 0}\frac{1}{t}\widetilde{\mathbb{P}}\left(\bar{U}_{1}^{+}\geq\frac{t^{-\frac{1}{Y}}x-G^{*}u}{M^{*}}\right)\!\right)du\nonumber\\
&=\frac{C}{Y}\left[M^{Y}-\left(M^{*}\right)^{Y}\right]x^{-Y}.
\end{align*}
For $I_{2}(t;w,x)$, since for any $x>0$ and $w>0$, ${\Red t^{-1/Y}}x-Mt^{-\frac{1}{Y}}w>0$ is equivalent to $t<w^{2}/(M^{2}w^{2})$,
\begin{align*}
0&\leq\frac{1}{t}\int_{\frac{t^{-\frac{1}{Y}}x-Mt^{\frac{1}{2}-\frac{1}{Y}}w}{M+G}}^{\frac{t^{-\frac{1}{Y}}x-M^{*}t^{\frac{1}{2}-\frac{1}{Y}}w}{M+G}}\widetilde{\mathbb{P}}\left(t^{\frac{1}{2}-\frac{1}{Y}}w+u\leq\bar{U}_{1}^{(p)}\leq\frac{t^{-\frac{1}{Y}}x-G^{*}u}{M^{*}}\right)p_{U}(1,u)\,du\nonumber\\
&\leq\frac{1}{t}\widetilde{\mathbb{P}}\left(\bar{U}_{1}^{(p)}\geq t^{\frac{1}{2}-\frac{1}{Y}}w\right)\widetilde{\mathbb{P}}\left(-\bar{U}_{1}^{(n)}\geq\frac{t^{-\frac{1}{Y}}x-Mt^{\frac{1}{2}-\frac{1}{Y}}w}{M+G}\right)\rightarrow 0,\quad t\rightarrow 0,
\end{align*}
which completes the proof.\hfill $\Box$

\bibliographystyle{plain}

\end{document}